\documentclass[a4paper,fleqn,usenatbib]{mnras}

\usepackage[T1]{fontenc}
\usepackage{ae,aecompl}

\usepackage{graphicx}	
\usepackage{amsmath}	
\usepackage{amssymb}	
\usepackage{textcomp}
\usepackage{hyperref}
\usepackage{breakurl}
\newcommand{\urlwofont}[1]{\urlstyle{same}\url{#1}}

\title[First results from GeMS/GSAOI for project SUNBIRD]{First results from GeMS/GSAOI for project SUNBIRD: Supernovae UNmasked By Infra-Red Detection}

\author[E. C. Kool et al.]{
E. C. Kool$^{1,2}$\thanks{E-mail: erik.kool@mq.edu.au},
S. Ryder$^{2}$,
E. Kankare$^{3}$,
S. Mattila$^{4}$,
T. Reynolds$^{4}$,
\newauthor
R. M. McDermid$^{1,2}$, 
M. A. P\'{e}rez-Torres$^{5}$,
R. Herrero-Illana$^{6}$,
M. Schirmer$^{7}$,
\newauthor
A. Efstathiou$^{8}$,
F. E. Bauer$^{9,10,11}$,
J. Kotilainen$^{12,4}$,
P. V\"ais\"anen$^{13}$,
\newauthor
C. Baldwin$^{1}$,
C. Romero-Ca\~{n}izales$^{10,14}$
and A. Alberdi$^{5}$
\\
$^{1}$Department of Physics and Astronomy, Macquarie University, NSW 2109, Australia\\
$^{2}$Australian Astronomical Observatory, 105 Delhi Rd, North Ryde, NSW 2113, Australia\\
$^{3}$Astrophysics Research Centre, School of Mathematics and Physics, Queen's University Belfast, Belfast BT7 1NN, UK\\
$^{4}$Tuorla observatory, Department of Physics and Astronomy, University of Turku, V\"ais\"al\"antie 20, FI-21500 Piikki\"o, Finland\\
$^{5}$Instituto de Astrof\'{i}sica de Andaluc\'{i}a (CSIC), Glorieta de la Astronom\'{i}a s/n, E-18080 Granada, Spain\\
$^{6}$European Southern Observatory (ESO), Alonso de C\'ordova 3107, Vitacura, Casilla 19001, Santiago de Chile, Chile\\
$^{7}$Gemini Observatory, Casilla 603, La Serena, Chile\\
$^{8}$School of Sciences, European University Cyprus, Diogenes Street, Engomi, 1516 Nicosia, Cyprus\\
$^{9}$Instituto de Astrof{\'{\i}}sica and Centro de Astroingenier{\'{\i}}a, Pontificia Universidad Cat{\'{o}}lica de Chile, Casilla 306, Santiago 22, Chile\\
$^{10}$Millennium Institute of Astrophysics (MAS), Nuncio Monse{\~{n}}or S{\'{o}}tero Sanz 100, Providencia, Santiago, Chile\\
$^{11}$Space Science Institute, 4750 Walnut Street, Suite 205, Boulder, Colorado 80301\\
$^{12}$Finnish Centre for Astronomy with ESO (FINCA), University of Turku, V\"ais\"al\"antie 20, FI-21500 Piikki\"o, Finland\\
$^{13}$South African Astronomical Observatory, P.O.Box 9, Observatory 7935, Cape Town, South Africa\\
$^{14}$N\'{u}cleo de Astronom\'{i}a de la Facultad de Ingenier\'{i}a y Ciencias, Universidad Diego Portales, Av. Ej\'{e}rcito 441, Santiago, Chile
}

\date{Accepted XXX. Received YYY; in original form ZZZ}

\pubyear{2017}

\begin{document}
\label{firstpage}
\pagerange{\pageref{firstpage}--\pageref{lastpage}}
\maketitle

\begin{abstract}
Core collapse supernova (CCSN) rates suffer from large uncertainties as many CCSNe exploding in regions of bright background emission and significant dust extinction remain unobserved. Such a shortfall is particularly prominent in luminous infrared galaxies (LIRGs), which have high star formation (and thus CCSN) rates and host bright and crowded nuclear regions, where large extinctions and reduced search detection efficiency likely lead to a significant fraction of CCSNe remaining undiscovered.
We present the first results of project SUNBIRD (Supernovae UNmasked By Infra-Red Detection), where we aim to uncover CCSNe that otherwise would remain hidden in the complex nuclear regions of LIRGs, and in this way improve the constraints on the fraction that is missed by optical seeing-limited surveys. We observe in the near-infrared 2.15 $\mu$m $K_s$-band, which is less affected by dust extinction compared to the optical, using the multi-conjugate adaptive optics imager GeMS/GSAOI on Gemini South, allowing us to achieve a spatial resolution that lets us probe close in to the nuclear regions. 
During our pilot program and subsequent first full year we have discovered three CCSNe and one candidate with projected nuclear offsets as small as 200 pc. 
When compared to the total sample of LIRG CCSNe discovered in the near-IR and optical, we show that our method is singularly effective in uncovering CCSNe in nuclear regions and we conclude that the majority of CCSNe exploding in LIRGs are not detected as a result of dust obscuration and poor spatial resolution.


\end{abstract}




\section{Introduction}


Luminous and ultraluminous infrared galaxies (LIRGs and ULIRGs; $L_{\text{IR}}$ > $10^{11}$ $L_{\odot}$ and $L_{\text{IR}}$ > $10^{12}$ $L_{\odot}$, respectively) exhibit high star formation (SF) rates, and their corresponding high infrared (IR) luminosity is due to UV light from hot young stars getting absorbed by dust and re-emitted in the IR. In the local Universe the relative SF contribution of (U)LIRGs is small, but by a redshift of $\sim$1 the SF in these dusty galaxies dominates over that in normal galaxies \citep{magnelli2011}.

Based on their $L_{\text{IR}}$, starburst dominated (U)LIRGs have an elevated expected core-collapse supernova (CCSN) rate of the order of one per year \citep{mattila2001}, which is $\sim$100 times larger than the Milky Way CCSN rate \citep{adams2013}. Due to the short lifetime of their progenitors, CCSNe act as a relatively direct tracer of the current rate of massive SF. However, even with the onset of wide field optical SN searches this past decade (such as CRTS, \citealt{drake2009}; ASAS-SN, \citealt{shappee2014}; Pan-STARRS1, \citealt{chambers2016}; SkyMapper, \citealt{scalzo2017}) surprisingly few CCSNe have been found in LIRGs.

This shortfall of CCSNe is not unique to LIRGs: \citet{horiuchi2011} claimed that up to half of all CCSNe beyond the local volume, as predicted by the well-defined SF rate history, are not observed: the so-called \emph{Supernova Rate Problem}. In the very local Universe, however, such a discrepancy is not seen \citep{botticella2012,xiao2015}, and \citet{cappellaro2015} argue that systematic errors in the SN and SF rates remain too large to invoke a supernova rate problem in the first place. One of the main uncertainties in the observed CCSN rate is the fraction that is missed in obscured and clumpy galaxies, such as LIRGs. Based on monitoring of the LIRG Arp 299, \citet{mattila2012} placed empirical limits on the fraction of CCSNe missed by optical surveys as a function of redshift, concluding that the optical missed-fraction rises from 20\% locally to 40\% at redshift of $\sim$1. These corrections were successfully applied to bridge the gap between observed and predicted CCSN rates in several studies \citep[e.g.,][]{dahlen2012, melinder2012}, or used as a high extinction scenario for CCSN rates at high redshift \citep[z=2;][]{strolger2015}, but are based on a small SN sample.

In order to improve the constraints on the missed fraction of CCSNe, a larger statistical sample of CCSNe in obscured and clumpy galaxies is required.
LIRGs fit these properties well and also display a clear deficit in CCSN discoveries. They have therefore been targets for several supernova surveys in the near-IR, where extinction is vastly reduced compared to optical wavelengths ($A_K \sim 0.1 \times A_V$), to uncover their CCSN population. \citet{mannucci2003} observed 46 local LIRGs in natural seeing conditions in \emph{K'}-band and detected three CCSNe, an order of magnitude smaller than the rate estimated from the galaxies $L_{\text{FIR}}$. \citet{cresci2007} observed 17 LIRGs with HST/NICMOS in the \emph{F160W} filter and did not find any confirmed SNe. \citet{miluzio2013} observed 30 LIRGs across three semesters in service time with HAWK-I on the VLT in natural seeing conditions in \emph{K}-band and detected five CCSNe. They claimed good agreement with the expected rate, but assumed a large fraction ($\sim$60-75\%) remained hidden in the nuclear regions (<2 kpc) due to reduced search efficiency and extinction.

The results of these studies demonstrate that there remains considerable uncertainty in the CCSN rate from LIRGs. With a dozen CCSN discoveries in total, the number of detections have lagged behind the expected CCSN rate, typically attributed to limitations in temporal coverage, lack of contrast against the extremely luminous background and/or inadequate spatial resolution in order to resolve the crowded and complex nuclear regions in LIRGs. Recent studies using near-IR ground-based adaptive optics (AO) imaging to provide the necessary high spatial resolution have had more success uncovering SNe in these conditions, with an additional five near-IR CCSN discoveries \citep{mattila2007, kankare2008, kankare2012}. Promisingly, several of these CCSNe have been within a few hundred pc from the hosts' nuclei, with extinctions up to 16 magnitudes in \emph{V}-band.

Building on these early results, we commenced in 2015 the SUNBIRD (Supernova UNmasked By Infra-Red Detection) project: a systematic search for CCSNe in a sample of LIRGs within 120 Mpc using laser guide star AO (LGSAO) imaging with the Gemini South Adaptive Optics Imager (GSAOI, \citealt{mcgregor2004, carrasco2012}) with the Gemini Multi-Conjugate Adaptive Optics System (GeMS, \citealt{rigaut2014, neichel2014a}) on the Gemini South telescope. The SUNBIRD project aims to characterise the population of CCSNe in the dusty and crowded star forming regions of LIRGs and in this way improve the constraints on the fraction of CCSNe missed due to dust obscuration and/or nuclear vicinity. In this first paper we introduce this ongoing survey and report on the first results of project SUNBIRD: three LIRG CCSN discoveries and one CCSN candidate. 

During preparation of this manuscript, project SUNBIRD has extended coverage to the Northern hemisphere through use of the Keck telescope. A description of the Keck campaign and a detailed CCSN rate analysis of the complete sample of LIRGs in SUNBIRD will appear in a forthcoming paper.

Section \ref{sec:project} of this paper contains a description of the survey, including the galaxy sample and observing strategy. In Section \ref{sec:red} we describe the data reduction and analysis. We present the SUNBIRD supernova detections made so far in Section \ref{sec:detections}, followed by Sections \ref{sec:results} and \ref{sec:discussion} where we investigate the supernova types and explore the impact of these new discoveries on the total number of CCSNe discovered in LIRGs, respectively. Finally in Section \ref{sec:conclusions} we draw our conclusions.
Throughout this paper we assume $H_0$ = 70 km s$^{-1}$ Mpc$^{-1}$, $\Omega_{\Lambda}$ = 0.7, and $\Omega_M$ = 0.3.

\begin{table*}
\centering
\label{tab:target_list}
\begin{tabular}{|l|c|c|r|c|l|c|} \hline
LIRG&RA&Dec.&Distance&log $L_{\textrm{IR}}$&r$_{CCSN}$&Epochs\\
&(J2000)&(J2000)&(Mpc)&$(L_{\odot})$&(yr$^{-1}$)&\#\\ \hline
NGC 1204		&03 04 40.5&-12 20 26	&64		&10.96&0.25				&1\\
ESO 491-G020	&07 09 47.0&-27 34 10	&43		&10.97&0.25				&1\\
MCG +02-20-003	&07 35 42.5&+11 42 36	&72		&11.13&0.37				&1\\
IRAS 08355-4944	&08 37 02.3&-49 54 32	&115	&11.62&1.12				&2\\
NGC 3110		&10 04 02.7&-06 28 35	&79		&11.36&0.14$^{\dagger}$	&4\\
ESO 264-G036	&10 43 07.0&-46 12 43	&99		&11.34&0.59				&4\\
ESO 264-G057	&10 59 02.4&-43 26 33	&82		&11.15&0.38				&1\\
NGC 3508		&11 03 00.1&-16 17 23	&61		&10.97&0.25				&2\\
ESO 440-IG058	&12 06 53.0&-31 57 08	&111	&11.45&0.51$^{\ast}$	&4\\
ESO 267-G030	&12 14 12.6&-47 13 37	&96		&11.26&0.49				&5\\
NGC 4575		&12 37 52.1&-40 32 20	&63		&11.03&0.29				&2\\
IRAS 17138-1017	&17 16 36.3&-10 20 40	&83		&11.49&0.75$^{\ast}$	&5\\
IRAS 18293-3413	&18 32 40.2&-34 11 26	&85		&11.74&1.97$^{\ast}$	&4\\ \hline
\end{tabular}
\caption{SUNBIRD GeMS/GSAOI LIRG sample. Distances are from the NASA/IPAC Extragalactic Database (NED\protect\footnotemark), Virgo/GA corrected. CCSN rates are based on the empirical relation with $L_{\textrm{IR}}$ from \citet{mattila2001}, unless otherwise indicated: CCSN rate based on SED fits from \citet{herrero2017}, denoted by $\ast$, or this work, denoted by $\dagger$. Values of log $L_{\textrm{IR}}$ are from \citet{sanders2003}, adjusted for updated distances. Final column shows number of epochs obtained with GeMS/GSAOI.}
\end{table*}
\section{Project SUNBIRD}
\label{sec:project}
\subsection{Galaxy sample}
\label{sec:sample}
The sample of LIRGs observed in SUNBIRD was selected from the IRAS Revised Bright Galaxy Sample \citep[RBGS;][]{sanders2003}. The main constraints to the sample selection originated from the instrument we used, GeMS/GSAOI on the Gemini South telescope, which requires guide stars of sufficient brightness and vicinity for the AO correction (see section~\ref{sec:strat}). Additionally we limited the sample to galaxies that are closer than 120 Mpc (z=0.027) in order to be able to resolve the central regions as close to the nucleus as possible. At this distance a typical AO corrected FWHM of 0.1\arcsec\ corresponds to $\sim$60 pc. We included targets with IR luminosities L(8-1000$\mu$m) in RBGS of log($L_{\text{IR}}$) > 10.9, see Table~\ref{tab:target_list}. Finally, we omit LIRGs where a significant AGN contamination to the IR luminosity could be expected, and thus exclude targets with `warm' IRAS colours, requiring f25/f60 < 0.2  \citep[e.g.][]{farrah2005}. The only exception is IRAS 08355-4944 with f25/f60 = 0.24 which is included in the sample based on SED fitting results, where it was shown to have a high SF rate of $\sim$85 M$_{\odot}$ yr$^{-1}$ \citep{dopita2011}, typical for a SF dominated LIRG. 

\subsection{Observing strategy}
\label{sec:strat}
The near-IR observations were obtained with GeMS/GSAOI on the Gemini South telescope. GSAOI is a near-IR AO imaging camera fed by GeMS and records images in a 85\arcsec $\times$ 85\arcsec\ field-of-view (FOV) with a pixel scale of 0.0197\arcsec\ pixel$^{-1}$, delivering close to diffraction limited images between 0.9 - 2.4 \textmu m. An optimal uniform AO correction across the FOV of GeMS requires three natural guide stars (NGS) in addition to the 5-point sodium laser guide star (LGS) pattern. The minimum requirement for AO correction at the time of the observations was at least one NGS of sufficient brightness ($m_\textrm{R}$ < 15.5 mag) available within the 1\arcmin\ patrol field of the wave front sensor probes and one on-detector guide window star ($m_\textrm{H}$ < 13.5 mag) within the 40\arcsec\ FOV of any of the four GSAOI detectors at all dither positions.

The SN search was conducted in $K_s$-band, as this is where, compared to \textit{J} and \textit{H}, AO performs best and extinction due to dust is lowest. Each target was imaged with a 9 step dither pattern for 120s at each position with a step size large enough (>5\arcsec) to cover the gaps between the detectors. The targets were typically centred on one of GSAOI's four arrays, with orientation depending on the galaxy and the locations of the NGS. Employing the efficient cadence strategy from \citet{mattila2001} for near-IR CCSN searches, we aimed to observe each galaxy twice each semester. In practice we achieved this cadence for half of the sample while the remainder of the sample galaxies was observed less frequently, due to seasonal weather variations, sodium layer return, and interruptions due to aircraft and satellite avoidance (see Table \ref{tab:target_list}). If a night did not allow for coverage of all observable targets, priority was given to galaxies with a high expected SN rate and those for which at least one GeMS/GSAOI epoch was already available. Expected SN rates were based on the empirical relation from \citep{mattila2001}:
\begin{equation}
r_{SN} = 2.7\times 10^{-12}\times L_{\text{IR}}/\text{L}_{\odot}\ \text{yr}^{-1}
\label{eq:snr}
\end{equation}

The targets in our sample with just one epoch were checked for SNe against archival high-resolution VLT/NACO (Nasmyth Adaptive Optics System Near-Infrared Imager and Spectrograph, 0.055\arcsec\ pixel$^{-1}$; \citealt{lenzen2003, rousset2003}) AO images, obtained by members of the SUNBIRD collaboration as part of a predecessor program \citep{randriamanakoto2013} and available for the whole sample.

Our total sample of LIRGs covered with GeMS/GSAOI so far consists of 13 galaxies up to a distance of 115 Mpc. Even though for some LIRGs there were only one or two NGS available and AO correction was not optimal, across our full data set a typical point-spread function (PSF) of $\sim$0.07\arcsec\ - 0.12\arcsec\ FWHM was achieved.
\footnotetext{The NASA/IPAC Extragalactic Database (NED) is operated by the Jet Propulsion Laboratory, California Institute of Technology, under contract with the National Aeronautics and Space Administration.}

\subsection{Multi-wavelength follow up}
Following a potential SN detection in $K_s$, the source was first checked for proper motion between exposures, to exclude a passing minor planet\footnote{\urlwofont{http://www.minorplanetcenter.net/cgi-bin/checkmp.cgi}}. 
Follow up with GeMS/GSAOI in \textit{H} and \textit{J} was done as soon as possible, which due to observing constraints typically occurred in the next GeMS/GSAOI observing window. As these observing windows were two to three months apart, rapid follow up of the SN candidate in the near-IR/optical was done with other instruments: in \textit{JHK} with NACO on the VLT, or contemporaneously with the Nordic Optical Telescope (NOT, \citealt{djupvik2010}) in \emph{r'}- and \emph{i'}-band with ALFOSC\footnote{The data presented here were obtained in part with ALFOSC, which is provided by the Instituto de Astrofisica de Andalucia (IAA-CSIC) under a joint agreement with the University of Copenhagen and NOTSA.} (Andalucia Faint Object Spectrograph and Camera, 0.19\arcsec\ pixel$^{-1}$) and in \textit{JHK} with NOTCam (Nordic Optical Telescope near-infrared Camera and spectrograph, 0.234\arcsec\ pixel$^{-1}$).

In addition to near-IR and optical imaging, two SN candidates were observed at radio wavelengths with the Karl G. Jansky Very Large Array (JVLA). A detection would provide important information about the nature of the SN and rule out a Type Ia SN, as even the most nearby Type Ia have yet to be detected at radio wavelengths \citep[e.g.][]{hancock2011,pereztorres2015,chomiuk2016}. The intrinsic rate of Type Ia SNe in LIRGS is estimated to be $\sim$5\%\ of that of CCSNe \citep{mattila2007}. As a final step, when possible near-IR spectroscopic coverage was obtained, under natural seeing conditions using the cross-dispersed mode of the Gemini Near Infra-Red Spectrograph (GNIRS) on Gemini North \citep{elias2006a,elias2006b}. The follow up observations are described in more detail in Section \ref{sec:detections}.

\section{Data reduction and analysis methods}
\label{sec:red}
\subsection{GeMS/GSAOI data reduction}
The GeMS/GSAOI data were reduced using \textsc{theli}\footnote{\urlwofont{https://www.astro.uni-bonn.de/theli/}} \citep{erben2005, schirmer2013}, generally following the procedures described in \citet{schirmer2013} and \citet{schirmer2015}. GeMS/GSAOI exposures suffer from a distortion pattern with a static component introduced by GeMS, and variable distortion components depending on NGS configurations, position angle and elevation \citep{neichel2014b, schirmer2015}. \textsc{theli} uses \emph{Scamp} \citep{bertin2006} for astrometric calibration and distortion correction of individual exposures prior to final co-addition, based on reference catalogues of point sources measured in distortion-free images of the same field, such as from HAWK-I on the VLT or VIRCAM on the VISTA telescope. In this way an optimal data quality across the FOV is obtained. Optimisations or adaptations of the procedures mentioned above are described in Appendix \ref{sec:theli_astro}. The solutions are specific to this project, but applicable to any GeMS/GSAOI observations with a low number source density.
\subsection{NOT and NACO data reduction}
The near-IR NOTCam instrument data were reduced with a slightly modified version of the external {\sc notcam} package\footnote{\urlwofont{http://www.not.iac.es/instruments/notcam/guide/observe.html}} (v. 2.5) within {\sc iraf}\footnote{\textsc{iraf} is distributed by the National Optical Astronomy Observatory, which is operated by the Association of Universities for Research in Astronomy (AURA) under cooperative agreement with the National Science Foundation \citep{iraf}}. The reduction steps included flat field correction, distortion correction, sky subtraction, and stacking of the individual exposures for increased signal-to-noise ratio. The optical \textit{r'}- and \textit{i'}-band ALFOSC images were reduced using the {\sc quba} pipeline \citep{valenti2011}, including bias subtraction and flat field correction.

The near-IR data taken with VLT/NACO were reduced using the NACO pipeline, which is based on the ESO Common Pipeline Library (CPL)\footnote{\urlwofont{http://www.eso.org/sci/software/cpl/}}. The jittered on-source images were flat field corrected, then median-combined to create a sky frame. Bad pixels were removed and the sky was subtracted from the individual images. The images were then stacked using a 2-D cross-correlation routine.

\subsection{GNIRS near-IR spectroscopy data reduction}
We used the cross-dispersed spectroscopy mode, providing a complete spectrum within 0.8-2.5~$\micron$  at an instrumental resolution of $R\sim 1700$. The data were reduced using version 2.6 of the XDGNIRS pipeline\footnote{\urlwofont{http://drforum.gemini.edu/topic/gnirs-xd-reduction-script/}}.
Briefly, the spectra were cleaned of pattern noise caused by the detector controller, using the python code from the Gemini website\footnote{\urlwofont{http://www.gemini.edu/sciops/instruments/gnirs/data-format-and-reduction/cleanir-removing-electronic-pattern-0}}. Radiation events from the radioactive lens coatings on the GNIRS short camera were identified and interpolated over using \textsc{iraf}'s {\it fixpix} task. Files were divided by a combined master flat field, created from a combination of quartz-halogen and IR flats taken after each science observation.
Subtraction of a sky frame removed night sky emission lines as well as other static artefacts in the detector, such as stable hot pixels. 
The orders were rectified using daytime pinhole flats, wavelength calibrated using an argon arc frame and 1D spectra were extracted from each order using \textsc{iraf} task {\it apall}. 
Telluric correction and flux calibration were done using an A-type star which was observed immediately before or after each science object. Each order of the science target was multiplied by a black body spectrum of the telluric star's effective temperature, scaled to the \emph{K}-band flux of the standard star for an approximate 2MASS flux calibration, and the orders were joined together using the \textsc{iraf} task {\it odcombine}.

\subsection{Image subtraction}
\label{sec:image_sub}
Subtraction of different epochs was done using a slightly modified (to accept manual stamp selection) version of image subtraction package ISIS 2.2 \citep{alard1998, alard2000}, where the software matches the PSF as well as flux and background levels of a previously aligned pair of images by deriving an optimal convolution kernel based on a selection of small windows (or ``stamps'') around objects with high signal-to-noise. See Figs. \ref{fig:iras18293}-\ref{fig:ngc3110} for the resulting subtractions. In the case of IRAS 18293-3413 the subtraction process resulted in considerable residuals at locations of high signal-to-noise, such as the nucleus and other bright compact sources. This was a result of having to prioritise an optimal subtraction for the central regions, as the SN is located very close to the nucleus. AO is optimized for point sources, but for a bright nuclear region with some intrinsic morphology the PSF is likely to vary between epochs. As such, the extraction of a SN signal in a location with such a steep background gradient is non-trivial, because any deviations from a perfect PSF match will produce large residuals. Despite our best efforts it was not possible to obtain a uniform subtraction across the full image. The smoothest subtraction of the nucleus was obtained by selecting a large number (>10) of small stamps near and in the galaxy to map the PSF around the SN location as well as possible, but this resulted in the residual patterns visible in the subtraction at the locations of bright compact objects in the field.

\subsection{Photometry}
The photometry of the objects was measured using the \textsc{snoopy}\footnote{\textsc{snoopy}, originally presented in \citet{patat}, has been implemented in \textsc{iraf} by E. Cappellaro. The package is based on {\it daophot}, but optimised for SN magnitude measurements} package in \textsc{iraf}, where a PSF is fitted to the SN residual in the subtracted image. The PSF was derived from 10 isolated field stars in the FOV of the one image that was not convolved during subtraction (i.e. the image with poorest image quality). \textsc{snoopy} removes a simple background estimate from the region surrounding the SNe (excluding the innermost region around the object) during PSF fitting, but as the SN detections are generally located in nuclear regions with large and variable background signal, the PSF fitting is performed on the SN residual in the subtracted image, where the complex background has already been removed. The photometry was calibrated against 5 2MASS stars in the FOV, when available. Systematic errors in the local background subtraction were estimated by simulating and PSF fitting nine artificial stars in a three by three grid pattern around the SN position in the subtracted image. These typically dominated the photometric errors.

The FOV of NGC 3110 did not offer any suitable catalogue reference stars and only one clear isolated field star for PSF fitting purposes. This field star acted as the PSF model and photometric reference for SN 2015ca after the field star's $JHK_s$ magnitudes were determined from NOT images calibrated against 5 2MASS sources. 

In case of a non-detection, a 5-$\sigma$ upper limit was determined by simulating stars of decreasing brightness at the SN location using the task {\it mkobjects} in \textsc{iraf} package {\it artdata}, prior to subtraction. The signal-to-noise was based on the aperture flux of the residual of the simulated star compared to 80 empty positions in the field around the SN location using \textsc{iraf} task {\it phot}. 

\subsection{Light curve template fitting}
To determine the type of the discovered SNe, the observed light curve data points were fitted using (reduced) $\chi^2$ minimization to three prototypical template light curves, representing Type IIP, IIn and stripped envelope Type IIb and Ib/c SNe, similarly to \citet{kankare2008,kankare2014}. The three prototypical CCSNe templates were chosen on the basis of being well characterized and having a well sampled light curve in the near-IR covering the different stages of SN evolution during at least the first 150 days since explosion. It should be noted that there is non-negligible diversity in SNe even within a given type, and thus the sources detected here could differ from the templates. The templates we used (see Fig. \ref{fig:2013if_fits}, \ref{fig:2015ca_fits} and \ref{fig:2015cb_fits}) are as follows:
\begin{itemize}
\item {\bf Type IIP:} A type IIP template fit was carried out based on the photometric evolution of SN 1999em. \textit{UBVRI} light curves of SN 1999em \citep{leonard2002} were transferred into \textit{ugri} using the conversions of \citet{jester2005}. \textit{JHK} light curves were obtained from \citet{krisciunas2009}. The distance modulus, total \textit{V}-band line-of-sight extinction, and explosion date of the used template SNe were adopted for the analysis from the literature. For SN 1999em, $\mu$ = 30.34 mag, $t_{\mathrm{e}}$ = 2451475.0 in JD, and $A_{\mathrm{V}}$ = 0.34 mag were reported by \citet{krisciunas2009}.\\

\item {\bf Type IIn:} A type IIn template fit was carried out using the published light curves of SN 1998S \citep{fassia2000,liu2000,mattila2001}. Similar to SN 1999em, the Johnson-Cousins light curves were converted into the SDSS system using the transformations of \citet{jester2005}. For SN 1998S, $\mu$ = 31.15 mag, $t_{\mathrm{e}}$ = 2450872.5 in JD, and $A_{\mathrm{V}}$ = 0.68 mag were adopted from \citet{fassia2000}.\\

\item {\bf Type IIb/Ib/Ic:} Well-sampled multiband light curves of Type IIb SN 2011dh \citep{ergon2014,ergon2015} were used as a general template for stripped-envelope SNe, since the photometric evolution of Type IIb SNe appears to be fairly similar to that of Type Ib/c SNe \citep[e.g.][]{arcavi2012}. \citet{ergon2014} finds for SN 2011dh values of $\mu$ = 29.46 mag, $t_{\mathrm{e}}$ = 2455713.0 in JD, and $A_{\mathrm{V}}$ = 0.22 mag.
\end{itemize}

All the bands are fitted simultaneously with three free parameters: the line-of-sight extinction $A_V$, time $t_0$ between explosion date and discovery, and a fixed constant \emph{C} applied to all bands, representing the intrinsic magnitude difference between SNe. Upper limits are used to constrain the template fits where necessary. Galactic line-of-sight extinction values are adopted from NED based on the dust maps of \citet{schlafly2011} and are fixed in the fit. For host galaxy and Galactic extinction, the \citet{cardelli1989} extinction law was used.

\begin{table}
\centering
\caption{Supernova near-IR and optical photometry}
\label{tab:mags}
\begin{tabular}{|l|c|c|l|} \hline
UT Date&Instrument&Filter&Magnitude \\ \hline
\multicolumn{4}{|c|}{\bf{SN 2013if (IRAS 18293-3413)}}\\ \hline
2012 September 14.1&NACO&$K_s$&>19.2\\
2013 April 21.3&GeMS/GSAOI&$K_s$&18.53$\pm$0.09\\
2013 May 8.4&NACO&$K_s$&18.61$\pm$0.12\\
2013 May 8.4&NACO&$H$&18.93$\pm$0.21\\
2013 May 24.3&GeMS/GSAOI&$K_s$&19.22$\pm$0.11\\
2013 June 11.1&GeMS/GSAOI&$K_s$&19.12$\pm$0.11\\
2013 June 11.1&GeMS/GSAOI&$H$&19.03$\pm$0.16\\
2013 June 11.1&GeMS/GSAOI&$J$&>18.7\\ \hline
\multicolumn{4}{|c|}{\bf{SN 2015ca (NGC 3110)}}\\ \hline
2015 March 11.1&GeMS/GSAOI&$K_s$&18.73$\pm$0.11\\
2015 March 27.8&NOT&$r'$&>20.0\\
2015 March 27.9&NOT&$i'$&>21.0\\
2015 April 5.9&NOT&$K_s$&17.91$\pm$0.22\\
2015 April 6.0&NOT&$H$&18.63$\pm$0.19\\
2015 April 6.0&NOT&$J$&19.39$\pm$0.21\\
2015 May 29.9&GeMS/GSAOI&$K_s$&20.22$\pm$0.12\\
2015 May 31.9&GeMS/GSAOI&$H$&20.67$\pm$0.09\\
2015 May 31.9&GeMS/GSAOI&$J$&21.25$\pm$0.08\\
2016 Feb. 19.2 &GeMS/GSAOI&$K_s$&>21.5\\
2016 Feb. 19.3 &GeMS/GSAOI&$H$&>22.0\\
2016 Feb. 19.4 &GeMS/GSAOI&$J$&>22.8\\ \hline
\multicolumn{4}{|c|}{\bf{SN 2015cb (IRAS 17138-1017)}}\\ \hline
2015 March 6.3&GeMS/GSAOI&$K_s$&16.56$\pm$0.09\\
2015 March 17.2&NOT&$r'$&>22.5\\
2015 March 17.3&NOT&$i'$&21.10$\pm$0.12\\
2015 April 6.2&NOT&$K_s$&17.12$\pm$0.22\\
2015 April 6.3&NOT&$H$&17.06$\pm$0.10\\
2015 April 6.3&NOT&$J$&18.34$\pm$0.18\\
2015 May 2.3&NOT&$J$&18.71$\pm$0.57\\
2015 June 1.2&GeMS/GSAOI&$K_s$&18.97$\pm$0.10\\ \hline
\multicolumn{4}{|c|}{\bf{AT 2015cf (NGC 3110)}}\\ \hline
2015 March 11.1&GeMS/GSAOI&$K_s$&20.97$\pm$0.13\\
2015 May 29.9&GeMS/GSAOI&$K_s$&21.4$\pm$0.2\\
2015 May 31.9&GeMS/GSAOI&$H$&>22.7\\
2015 May 31.9&GeMS/GSAOI&$J$&>22.8\\ \hline 
\end{tabular}
\end{table}
\section{Observations}
\label{sec:detections}
\subsection{SN 2013if in IRAS 18293-3413}
\begin{figure*}
	\includegraphics[trim=20 20 10 10, clip, width=\textwidth]{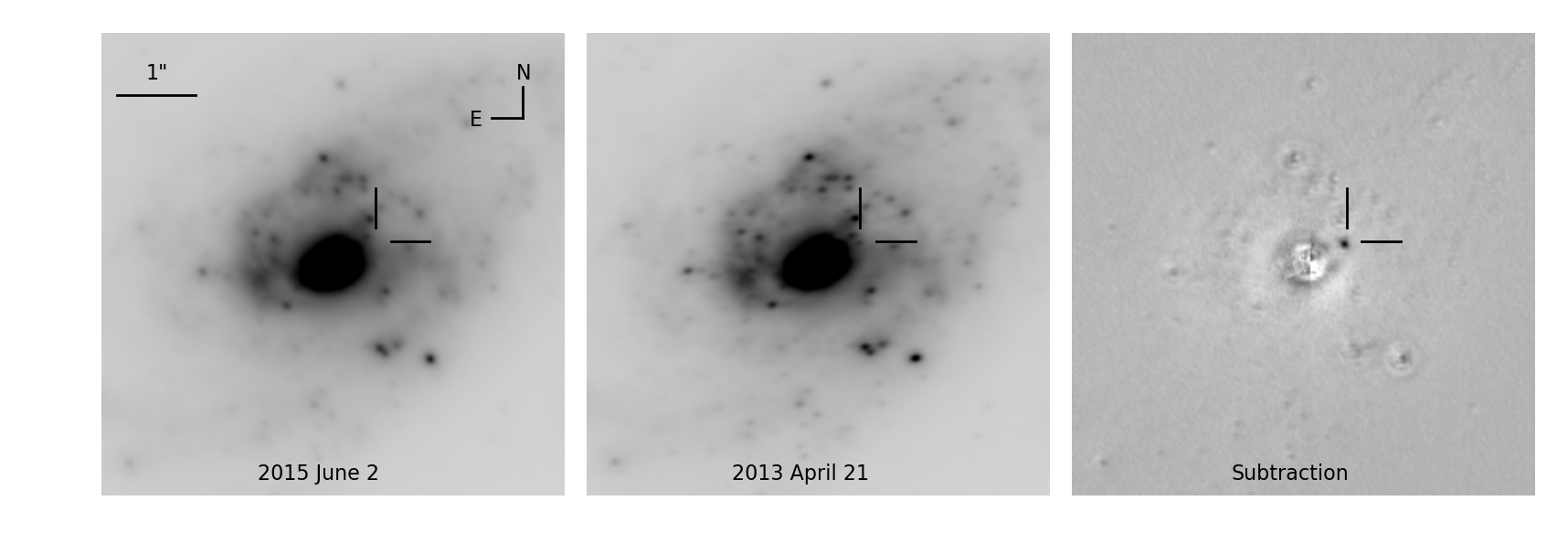}
    \caption{SN 2013if at R.A. = $18^{\textrm{h}}32^{\textrm{m}}41.10^{\textrm{s}}$ and Decl. = $-34^\circ 11\arcmin 27.24\arcsec$ in IRAS 18293-3413 with GeMS/GSAOI. From left to right with linear scaling: Reference image (June 2015), discovery image (April 2013) and the image subtraction.}
    \label{fig:iras18293}
\end{figure*}

SN 2013if \citep{2013if} in IRAS 18293-3413 was discovered with GeMS/GSAOI on 2013 April 21, see Fig.~\ref{fig:iras18293}. Subtraction of a $K_s$-band image taken with NACO on the VLT on 2004 September 13 \citep{mattila2007} showed a positive residual 0.2\arcsec\ North and 0.4\arcsec\ West (200 pc projected distance) from the nucleus. WCS matching in \textsc{theli} with a catalogue of 180 sources extracted from a VISTA image (VISTA Hemisphere Survey or VHS\footnote{\urlwofont{http://www.vista-vhs.org/}}) yielded R.A. = $18^{\textrm{h}}32^{\textrm{m}}41.10^{\textrm{s}}$ and Decl. = $-34^\circ 11\arcmin 27.24\arcsec$, with 0.03\arcsec\ and 0.03\arcsec\ uncertainty in R.A. and Decl., respectively. SN positions in this paper were determined using centroiding in \textsc{iraf} in the subtracted images to avoid the effects of strong background. Follow-up observations were made with NACO on 2013 May 8 in $K_s$ and $H$, and with GeMS/GSAOI in $K_s$ on 2013 May 24 and in $K_s$, \emph{H} and \emph{J} on 2013 June 11. SN 2013if was detected in all the follow up images with the exception of the final \emph{J}-band image, see Table~\ref{tab:mags}. Supernova free comparison GeMS/GSAOI images required for optimal image subtraction  were obtained in $K_s$, \emph{H} and \emph{J} on 2015 June 2. 

\begin{figure*}
	\includegraphics[trim=20 20 16 10, clip, width=\textwidth]{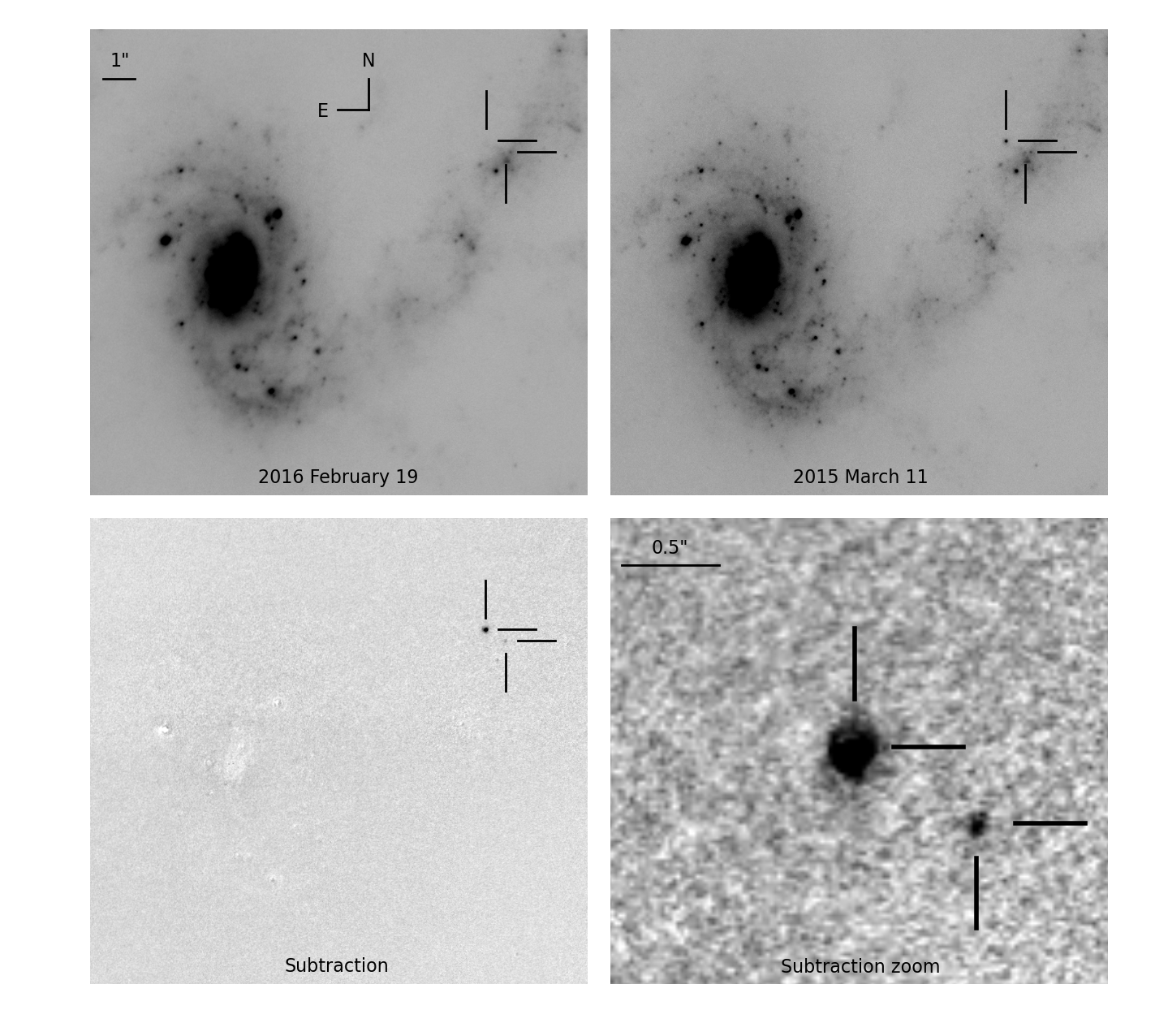}
    \caption{SN 2015ca and AT 2015cf in NGC 3110 with GeMS/GSAOI, at R.A. = $10^{\textrm{h}}04^{\textrm{m}}01.57^{\textrm{s}}$ and Decl. = $-06^\circ 28\arcmin 25.48\arcsec$ and R.A. = $10^{\textrm{h}}04^{\textrm{m}}01.53^{\textrm{s}}$ and Decl. = $-06^\circ 28\arcmin 25.84\arcsec$, respectively. Top row, with linear scaling, shows the reference image (February 2016) and discovery image (March 2015). Bottom row shows the full image subtraction and zoomed in around SN 2015ca, which shows AT 2015cf visible to the South-West.}
    \label{fig:ngc3110}
\end{figure*}

\begin{figure*}
	\includegraphics[trim=13 20 7 10, clip, width=\textwidth]{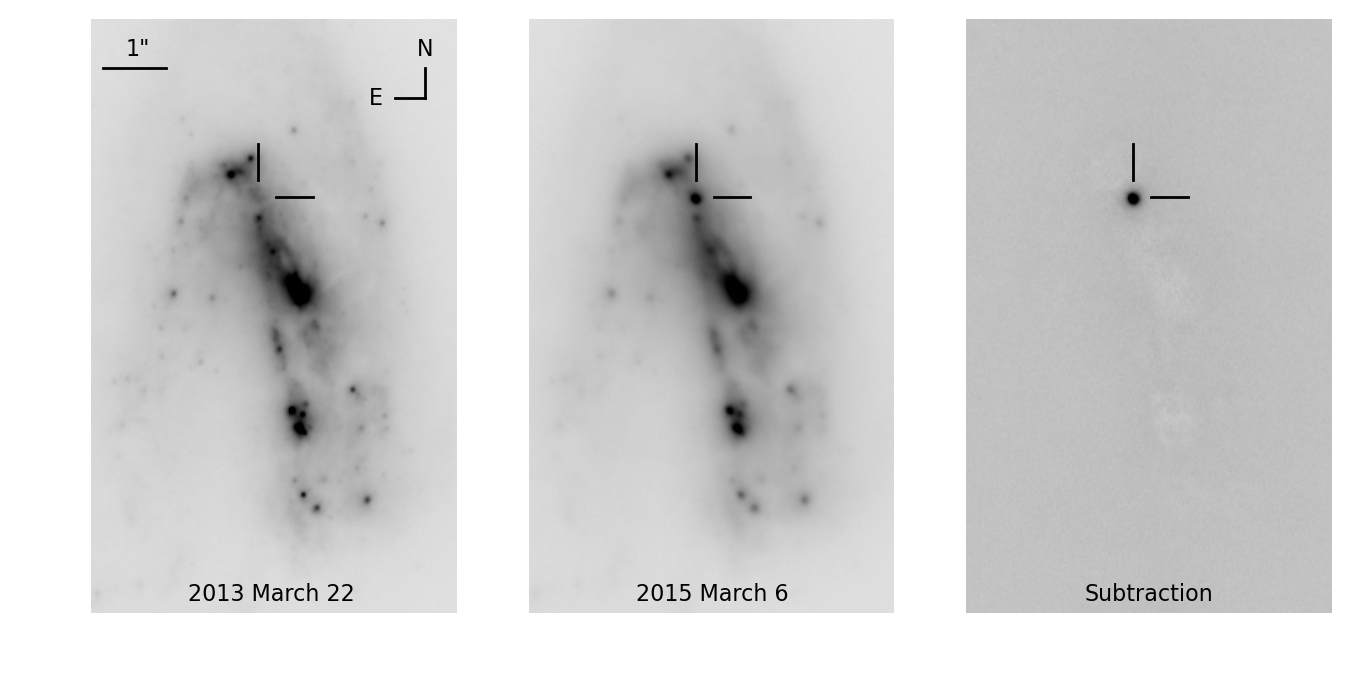}
    \caption{SN 2015cb at R.A. = $17^{\textrm{h}}16^{\textrm{m}}35.84^{\textrm{s}}$ and Decl. = $-10^\circ 20\arcmin 37.48\arcsec$ in IRAS 17138-1017 with GeMS/GSAOI. From left to right, with linear scaling: Reference image (March 2013), discovery image (March 2015) and image subtraction.}
    \label{fig:iras17138}
\end{figure*}
\subsection{SN 2015ca in NGC 3110}
The discovery image of SN 2015ca \citep{2015ca} in NGC 3110 was observed on 2015 March 11; see Fig.~\ref{fig:ngc3110}. As a reference image a NACO $K_s$-band image from 2010 December 28 was used \citep{randriamanakoto2013}. The subtracted image revealed a point source along the northern spiral arm of the galaxy 4.3\arcsec\ North and 8.0\arcsec\ West from the nucleus, corresponding to a projected distance of $\sim$3.5 kpc. WCS matching in \textsc{theli} with a catalogue of >300 sources extracted from a HAWK-I image \citep{miluzio2013} yielded R.A. = $10^{\textrm{h}}04^{\textrm{m}}01.57^{\textrm{s}}$ and Decl. = $-06^\circ 28\arcmin 25.48\arcsec$, with 0.03\arcsec\ and 0.04\arcsec\ uncertainty in R.A. and Decl., respectively. Follow up observations with the NOT were carried out on 2015 March 27 in $r'$ and $i'$ and on April 5 in $K_s$, $H$ and $J$, and with GeMS/GSAOI in $K_s$, $H$ and $J$ on 2015 May 29 and 31 and in $K_s$, $H$ and $J$ on 2016 February 19. SN 2015ca was detected in all near-IR bands in April and May 2015, but was not detected in any optical bands; see Table~\ref{tab:mags}. It had faded below our detection limits in February 2016.
\subsection{SN 2015cb in IRAS 17138-1017}
SN 2015cb \citep{2015cb} in IRAS 17138-1017 was discovered with GeMS/GSAOI on 2015 March 6; see Fig.~\ref{fig:iras17138}. Subtraction of a GeMS/GSAOI image from 2013 March 22 revealed a residual point source 1.4\arcsec\ North and 0.6\arcsec\ East (600 pc projected distance) from the nucleus. WCS matching in \textsc{theli} with a catalogue of 76 sources extracted from a VISTA image yielded R.A. = $17^{\textrm{h}}16^{\textrm{m}}35.84^{\textrm{s}}$ and Decl. = $-10^\circ 20\arcmin 37.48\arcsec$, with 0.04\arcsec\ and 0.04\arcsec\ uncertainty in R.A. and Decl., respectively. Follow up observations with the NOT were carried out on 2015 March 17 in optical (in \emph{i'}- and \emph{r'}-band, FWHM $\sim$1.1\arcsec), on April 6 in $K_s$, $H$ and $J$ (FWHM $\sim$1.0\arcsec), and with GeMS/GSAOI in $K_s$ on 2015 June 1. SN 2015cb was detected in all follow up observations except in \emph{r'}; see Table~\ref{tab:mags}.
\subsection{AT 2015cf in NGC 3110}
\label{littlebro}
The discovery image of SN 2015ca in NGC 3110 from 2015 March 11 showed a second transient source at R.A. = $10^{\textrm{h}}04^{\textrm{m}}01.53^{\textrm{s}}$ and Decl. = $-06^\circ 28\arcmin 25.84\arcsec$; see Fig.~\ref{fig:ngc3110}. AT 2015cf \citep{2015cf} is just 0.6\arcsec\ S and 0.4\arcsec\ W of SN 2015ca, as shown in the zoomed panel in Fig. \ref{fig:ngc3110}. The PSF of this source matches well with that of SN 2015ca and field stars in the image. The magnitude for this source at this epoch, bootstrapped off of SN 2015ca, is 20.97$\pm$0.13. It is not visible in any of the NOT epochs, which is not surprising as with a mere 0.7\arcsec\ separation from SN 2015ca it was likely blended with it in the subtraction. The GeMS/GSAOI $K_s$ image from 2015 May 29 does show a residual at the same position with a magnitude of 21.4$\pm$0.2, confirming that it is in fact a real transient. The $H$ and $J$ observations from the same epoch did not show the source, likely as a result of poorer image quality and/or due to significant extinction. In February 2016 it was not visible in any band.
\subsection{Radio observations}

\subsubsection{VLA observations}
We observed IRAS~17138-1017 and NGC~3110 with the VLA on 2015 April 8-9 under program 15A-471 (PI: P\'erez-Torres), while the VLA was in B-configuration.  We observed IRAS~17138-1017 at K-band (centred at 22 GHz) with a total bandwidth of 8 GHz, and NGC~3110 at X-band (10 GHz), with 4 GHz of bandwidth, in both cases using full polarization.

We used the bright quasar 3C286 for flux and bandpass calibration, and J0943-0819 and J1733-1304 to calibrate the phases of IRAS~17138-1017 and NGC~3110, respectively.
We performed a standard data reduction using the Common Astronomy Software Applications package \citep[CASA;][]{mcmullin2007}. We imaged our datasets using multi-frequency synthesis (MFS) with natural weighting, yielding a synthesized beam size and an rms of 0.56\arcsec $\times$ 0.30\arcsec and 9 $\mu\mathrm{Jy}\,\mathrm{beam}^{-1}$ for IRAS~17138-1017 and 0.96\arcsec $\times$ 0.66\arcsec and 15 $\mu\mathrm{Jy}\,\mathrm{beam}^{-1}$ for NGC~3110, respectively.

\begin{figure}
	\includegraphics[width=\columnwidth]{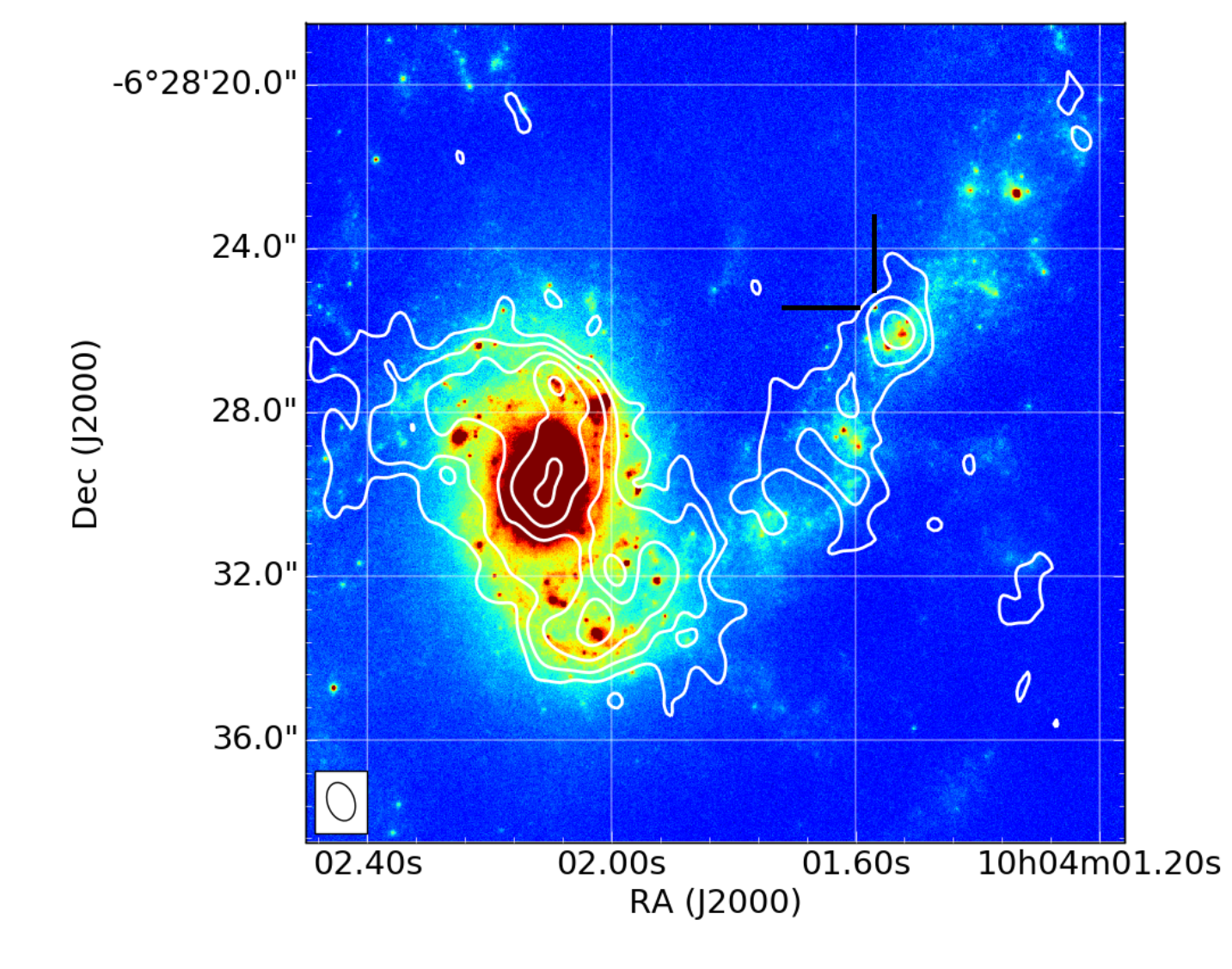}
    \caption{X-band (10\,GHz) VLA contours overlaid on GeMS/GSAOI detection image of SN 2015ca and AT 2015cf. SN 2015ca is indicated by tick marks, AT 2015cf is separated by just 0.7\arcsec.}
    \label{fig:ngc3110_vla}
\end{figure}

\begin{figure}
	\includegraphics[width=\columnwidth]{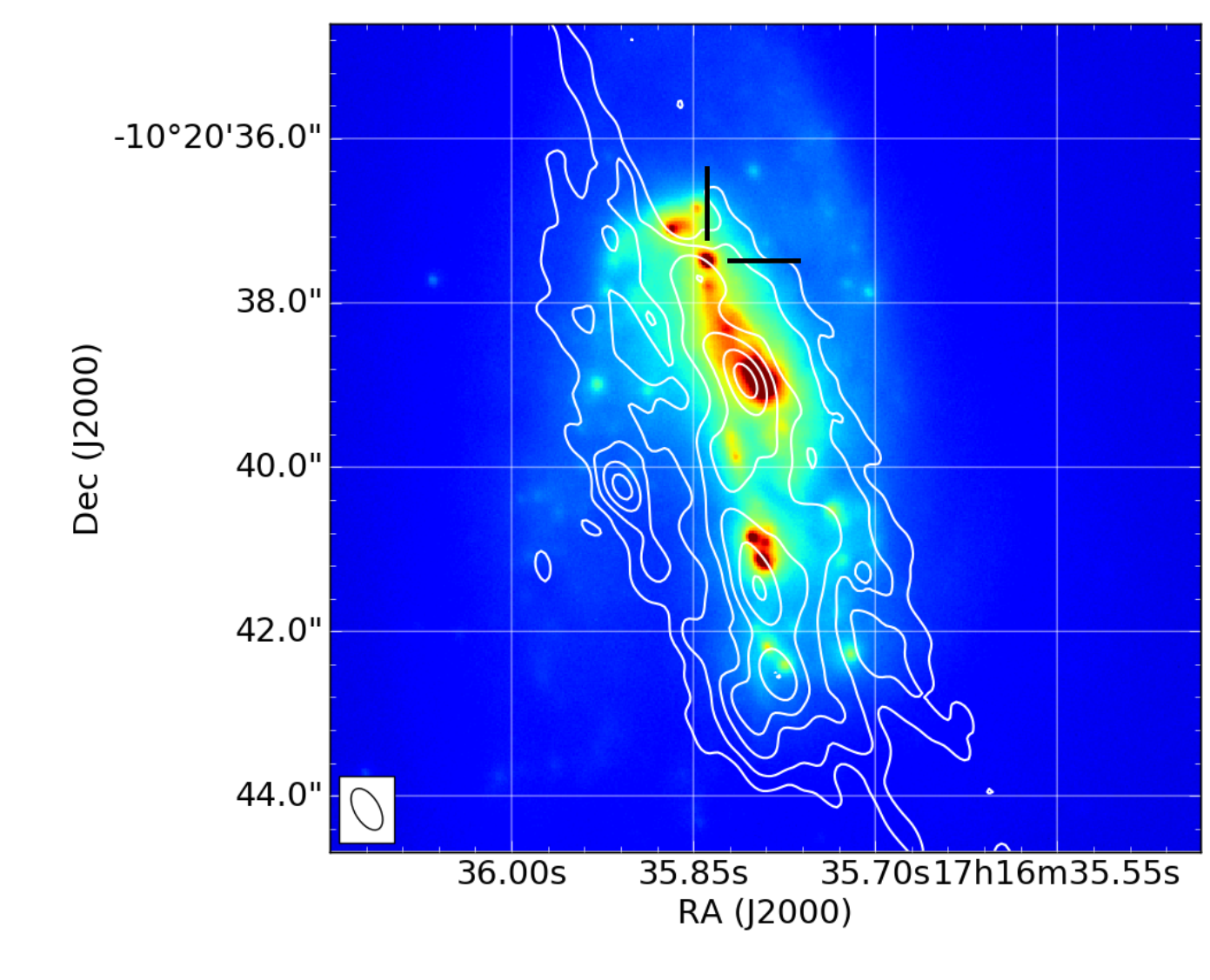}
    \caption{K-band (22\,GHz) VLA contours overlaid on GeMS/GSAOI detection image of SN 2015cb, with the SN indicated by tick marks.}
    \label{fig:iras17138_vla}
\end{figure}

Neither observation showed a local maximum coincident with the SN position. In the case of SN 2015cb in IRAS 17138-1017, any point source was most likely blended with the significant background signal in the central regions of the LIRG; see Fig.~\ref{fig:iras17138_vla}. SN 2015ca and AT 2015cf in NGC 3110 also coincide with local radio emission potentially masking possible source detections; see Fig.~\ref{fig:ngc3110_vla}. For further details on the radio limits, see Section \ref{sec:results}.

\subsubsection{eEVN observation}
We also observed the region around SN 2015ca and AT 2015cf with the electronic European VLBI Network (eEVN) on 10 May 2016. We used the eEVN at an observing frequency of 5.0 GHz, using an array of nine telescopes for about 2-hr, including overheads for calibration purposes and slew time.  
We observed the transients phase-referenced to the nearby source J0959-0828, 
using a typical duty cycle of four minutes. We used the strong source 3C273B as fringe finder and bandpass calibrator. All the data were correlated at the EVN MkIV data processor of the Joint Institute for VLBI in Europe (JIVE, the Netherlands), using an averaging time of 2 s.

We used AIPS for calibration, data inspection, and flagging of our eEVN data, using standard procedures. 
We then imaged a FOV of 1\arcsec$\times$1\arcsec\ centred at RA=10:04:01.572 and DEC=-06:28:25.480, and applied standard imaging procedures using AIPS, without averaging the data either in time, or frequency, to prevent time- and band- width smearing of the images.  
We did not detect any  signal above 101 $\mu$Jy/b (=$5\sigma$) in the field surrounding SN 2015ca. Several sources at the 3-$\sigma$ level were detected, but repeated imaging with different cleaning schemes showed that these were spurious detections, implying that there was no evidence of radio emission from SN 2015ca. AT 2015cf was not covered in the 1\arcsec$\times$1\arcsec\ FOV.

\subsection{Near-IR spectroscopic observations}
We obtained near-IR spectroscopy of SN 2015ca and SN 2015cb through Director's Discretionary Time with GNIRS on the Gemini North telescope on 2015 May 23. These observations were obtained in natural seeing conditions $\sim$70 days after discovery. We opted for near-IR spectroscopy as the SNe were located in crowded regions and expected to be significantly dust extincted. Image quality in the near-IR is typically better and less affected by dust extinction than the optical. The SNe were too faint to acquire directly, so a blind offset from a bright star was required. There was no clear indication of a point source in the slit, and the spectra did not show any emission lines associated with CCSNe. As such the obtained spectra could not be used to constrain the SN types of SN 2015ca or SN 2015cb.

\section{Analysis}
\label{sec:results}
\begin{table}
\label{tab:template}
\begin{tabular}{|l|c|c|c|c|} \hline
Template&$A_V$&$t_0$&$C$&$\tilde{\chi}^2$ \\ \hline
\multicolumn{5}{|c|}{\bf{SN 2013if} (IRAS 18293-3413)}\\ \hline
IIP, plateau&$0.0^{+2.8}_{-0.0}$&$67^{+1}_{-14}$&$1.8^{+0.1}_{-0.4}$&7.1\\[5pt] 
IIP, tail&$0.0^{+2.5}_{-0.0}$&$136^{+83}_{-13}$&$0.0^{+0.5}_{-1.7}$&4.7\\ [5pt]
IIn&$0.0^{+2.7}_{-0.0}$&$19^{+12}_{-3}$&$3.1^{+0.1}_{-0.5}$&5.0\\[5pt] 
IIb/Ib/Ic&$0.0^{+2.9}_{-0.0}$&$18^{+3}_{-2}$&$1.5^{+0.1}_{-0.5}$&7.4\\ \hline
\multicolumn{5}{|c|}{\bf{SN 2015ca} (NGC 3110)}\\ \hline
IIP&$3.4^{+1.0}_{-1.7}$&$58^{+14}_{-5}$&$1.5^{+0.4}_{-0.3}$&3.8\\[5pt]
IIn&$2.7^{+0.2}_{-0.7}$&$25^{+17}_{-9}$&$3.0^{+0.2}_{-0.2}$&10.3\\[5pt] 
IIb/Ib/Ic&$2.8^{+0.3}_{-0.4}$&$16^{+5}_{-2}$&$1.3^{+0.2}_{-0.2}$&5.0\\[5pt] 
Ia&$6.9^{+0.1}_{-0.7}$&$5^{+1}_{-1}$&$0.4^{+0.3}_{-0.2}$&17.9\\ \hline
\multicolumn{5}{|c|}{\bf{SN 2015cb} (IRAS 17138-1017)}\\ \hline
IIP, plateau&$4.6^{+0.3}_{-0.1}$&$64^{+11}_{-7}$&$-0.7^{+0.1}_{-0.3}$&11.7\\[5pt] 
IIP, tail&$3.6^{+0.4}_{-0.1}$&$134^{+15}_{-4}$&$-2.2^{+0.2}_{-0.3}$&29.0\\[5pt]
IIn&$5.2^{+0.3}_{-0.1}$&$24^{+5}_{-4}$&$0.5^{+0.2}_{-0.2}$&32.4\\[5pt] 
IIb/Ib/Ic&$4.7^{+0.1}_{-0.6}$&$19^{+10}_{-1}$&$-1.1^{+0.1}_{-0.2}$&11.2\\ \hline
\multicolumn{5}{|c|}{\bf{AT 2015cf} (NGC 3110)}\\ \hline
IIP&> 7.0&> 139&< 1.6&-\\[5pt]
IIb/Ib/Ic&> 0.0&> 105&< 2.0&-\\ \hline
\end{tabular}
\caption{Results from fitting light curve templates to the three SNe, with line-of-sight extinction $A_V$, time $t_0$ between explosion date and discovery, and a fixed constant \emph{C} representing the intrinsic magnitude difference between SNe. The final column shows the resulting $\tilde{\chi}^2$ for each fit.}
\end{table}

\begin{table}
\centering
\begin{tabular}{|l|c|} \hline
Total luminosity ($10^{11} L_{\odot}$)&$2.59^{+0.03}_{-0.08}$\\[5pt]
Starburst luminosity ($10^{11} L_{\odot}$)&$0.42^{+0.08}_{-0.02}$\\[5pt]
Disk luminosity ($10^{11} L_{\odot}$)&$2.17^{+0.03}_{-0.13}$\\[5pt]
AGN luminosity ($10^{11} L_{\odot}$)&<0.01\\[5pt]
SF rate, averaged&\\
over the past 50 Myr ($M_{\odot}~\mathrm{yr}^{-1}$)&$12.7^{+0.7}_{-1.1}$\\[5pt]
Starburst age (Myr)&$13.8^{+0.9}_{-2.4}$ \\[5pt]
Core-collapse supernova rate ($\mathrm{SN~yr}^{-1}$)&$0.14^{+0.01}_{-0.01}$\\ \hline
\end{tabular}
\caption{NGC 3110 SED model fitting parameters and derived physical quantities}
\label{tab:sed}
\end{table}
\subsection{SN 2013if}
\begin{figure*}
\centering
	\begin{tabular}{ccccc}
        \includegraphics[width=0.45\textwidth]{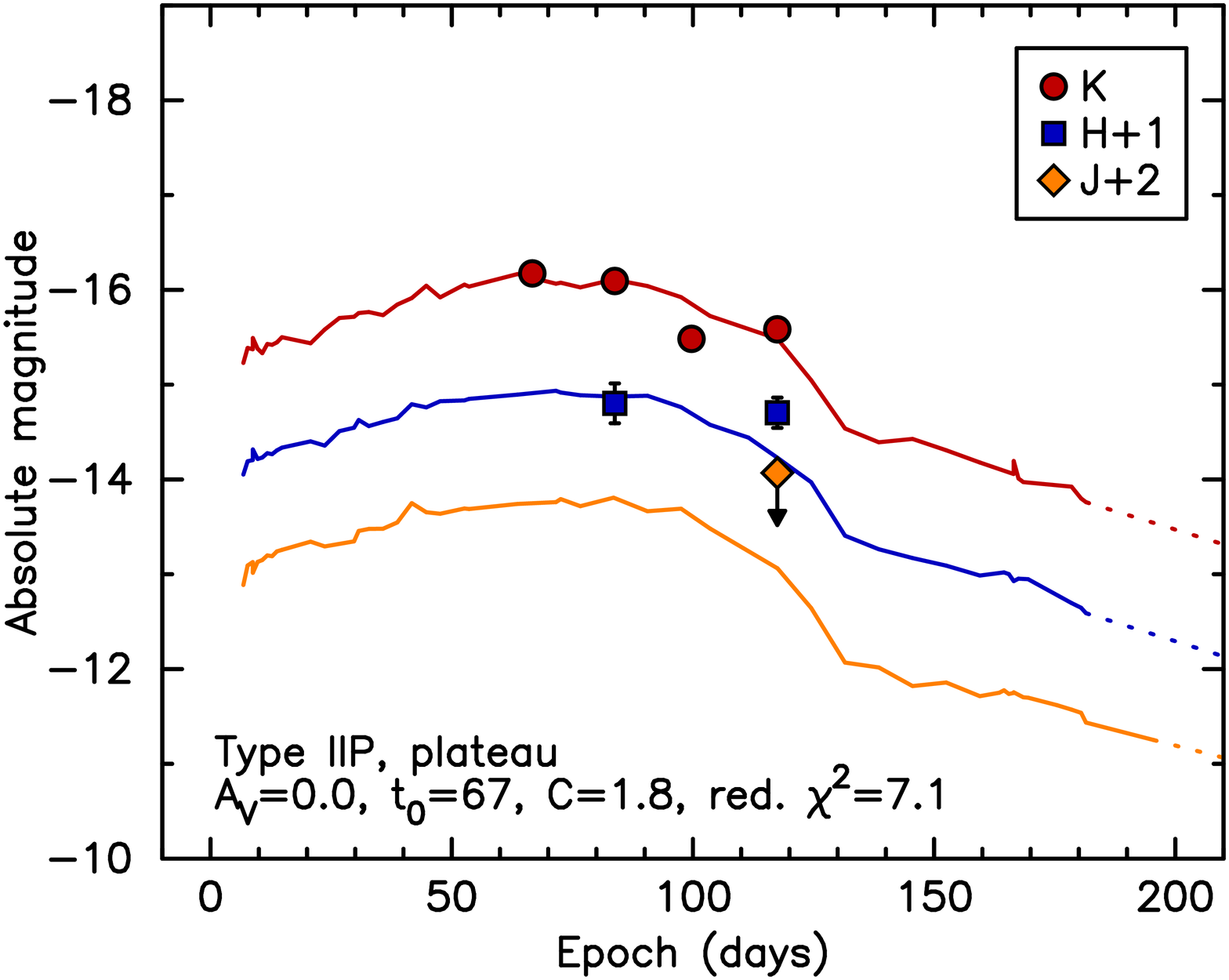} &
        \includegraphics[width=0.45\textwidth]{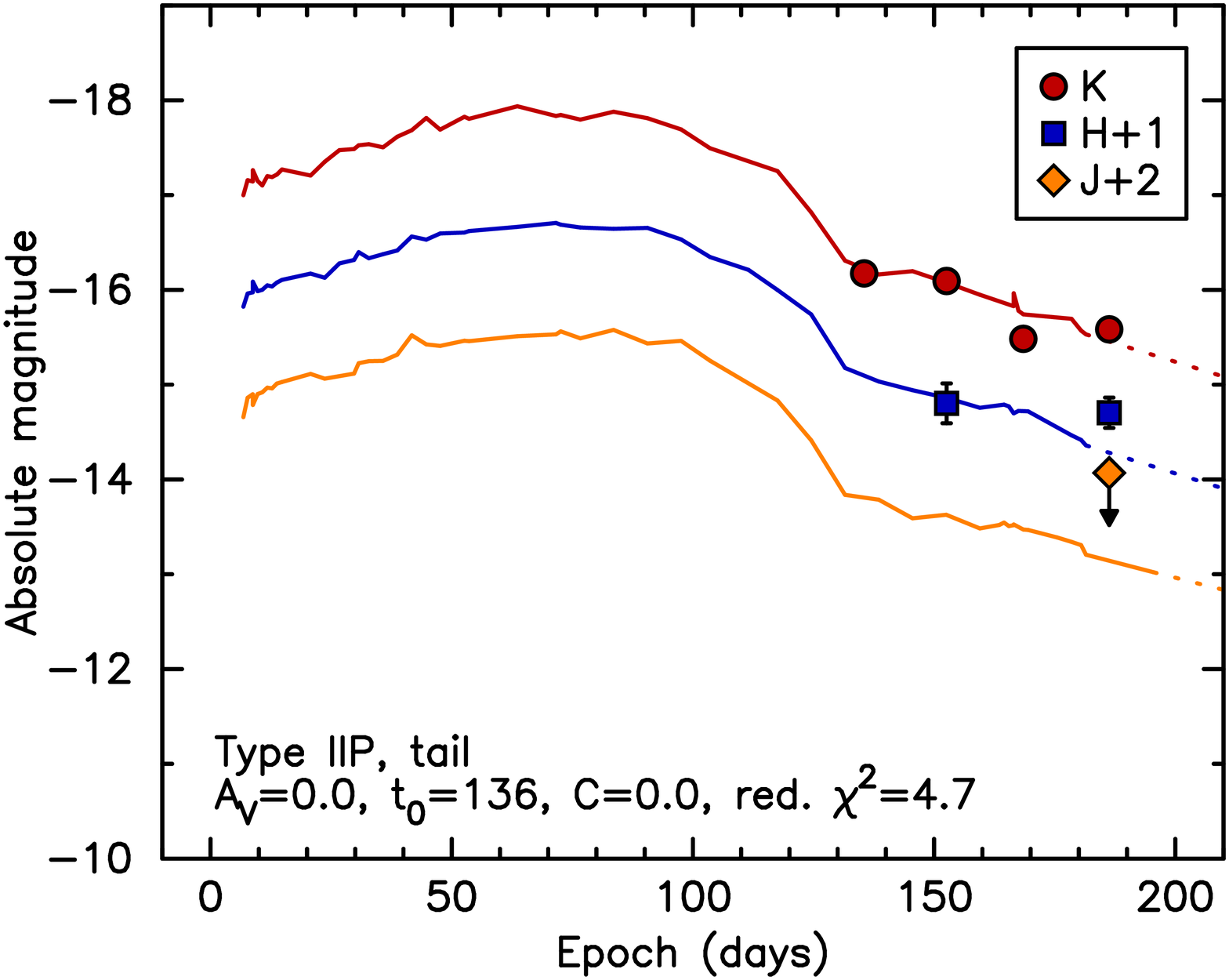} \\
        \includegraphics[width=0.45\textwidth]{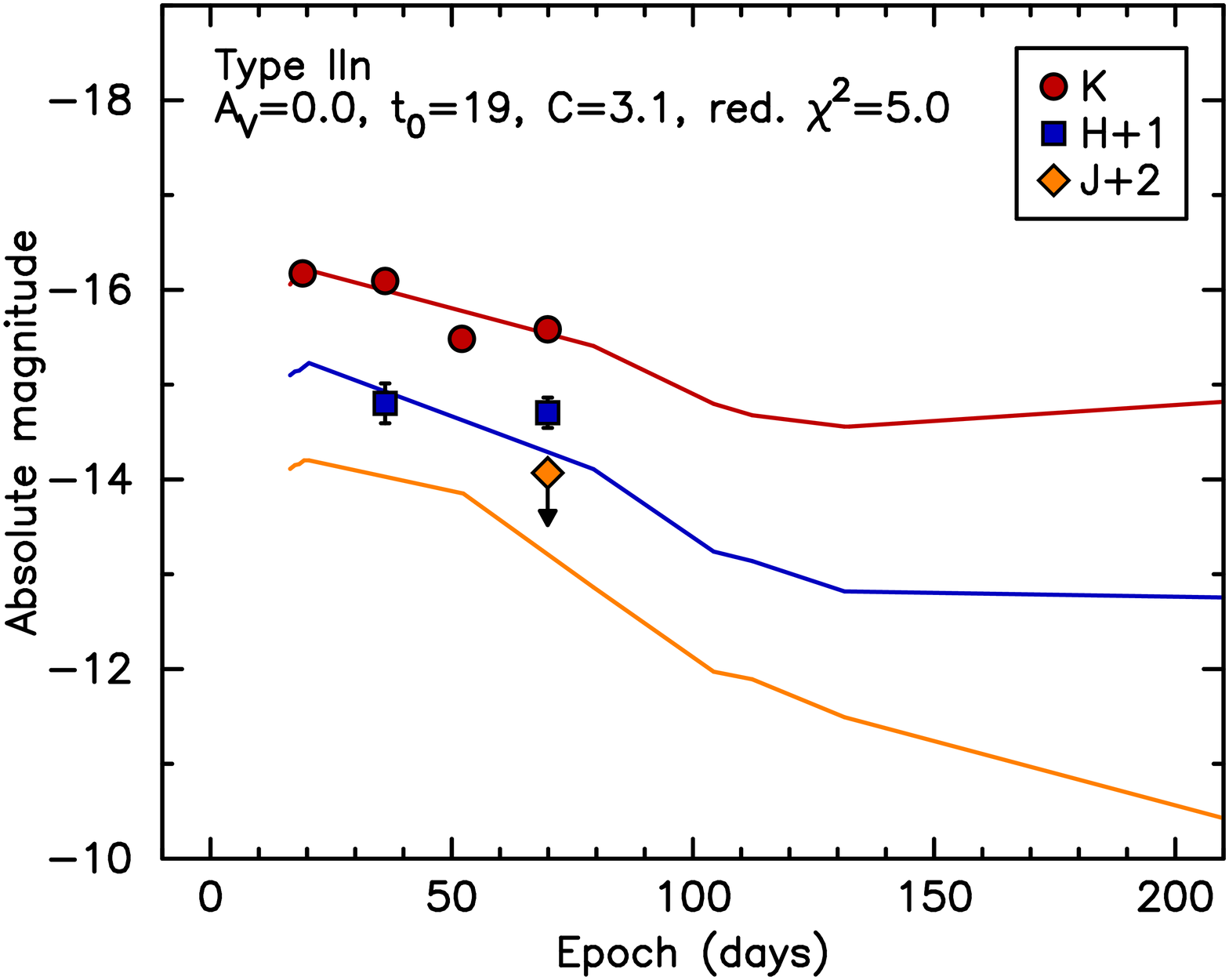} &
        \includegraphics[width=0.45\textwidth]{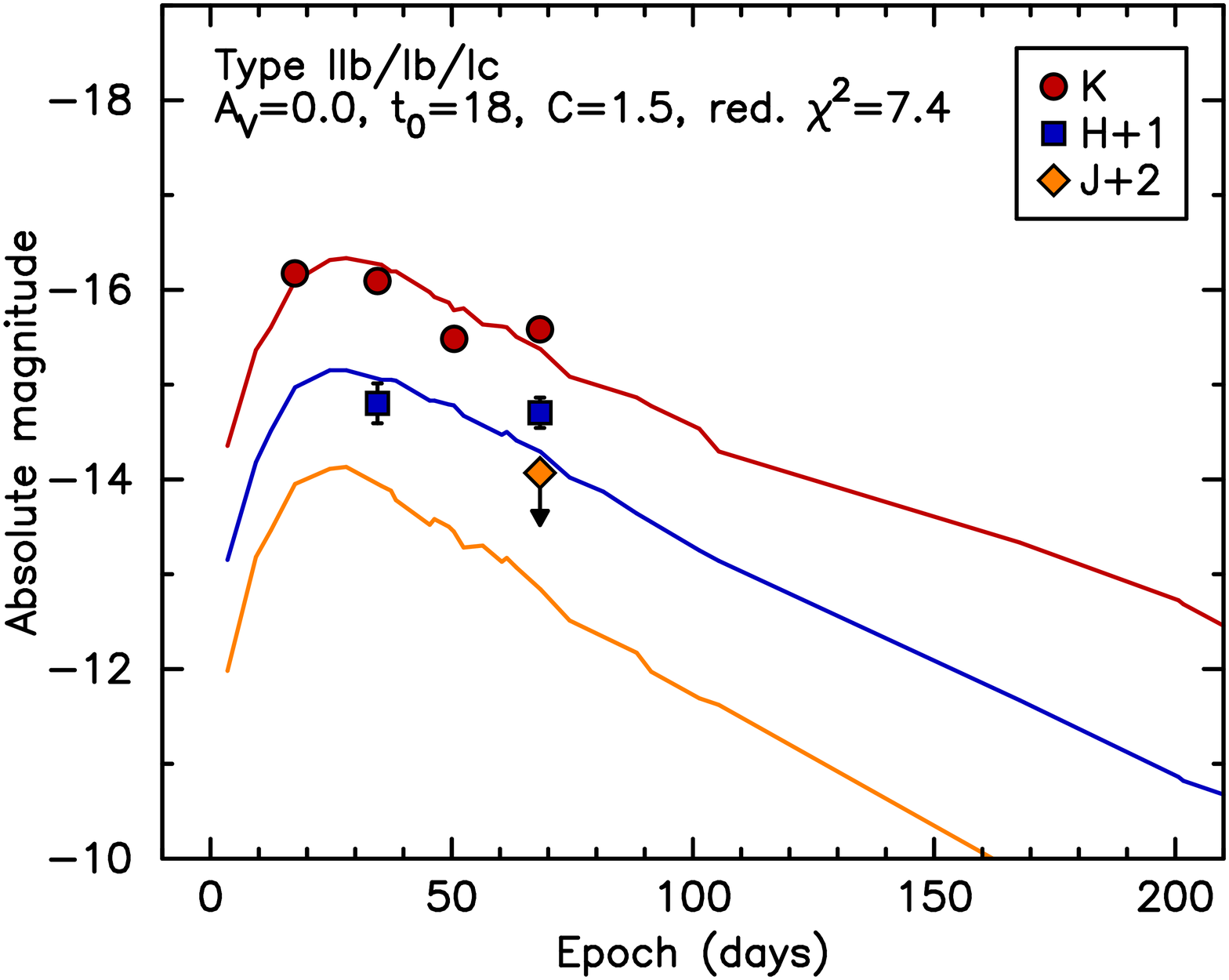} \\
	\end{tabular}
    \caption{Template light curve fits for SN 2013if in IRAS 18293-3413 are shown: Type IIP fitted in the plateau and tail phase, Type IIn, and Type IIb/Ib/Ic}
\label{fig:2013if_fits}
\end{figure*}

The best template fit for SN 2013if with a $\tilde{\chi}^2$ of 4.7 is a Type IIP caught in the tail phase (see Fig.~\ref{fig:2013if_fits} and Table~\ref{tab:template}), although the other templates fit the data with comparable values of $\tilde{\chi}^2$. A Type IIP caught in the tail phase is deemed most likely as, in addition to the fitting results, it is the only fit where no magnitude shift \emph{C} is required and the SN discovery is not before or at the maximum of the light curve. This means it does not require a sub-luminous supernova as opposed to the Type IIn fit with similar $\tilde{\chi}^2$, and it does not require discovery during the very short-lived phase around maximum light. 

The fit of the tail phase Type IIP is along a linearly extrapolated tail with a very slow colour evolution, so the fitting parameter time $t_0$ between explosion and detection is poorly constrained by the data. However a NACO image from 2012 September 14, 219 days prior to discovery, does not show a detection with an upper limit of 19.2 in $K_s$, allowing for the constraints on $t_0$. Regardless of SN type, it is noteworthy that all four template fits are best fitted with an extinction $A_V$ of 0. This is surprising as the SN is very close to the nucleus where significant extinction would be expected. Negligible extinction suggests we are observing the SN in the foreground of the host's nuclear regions.

\subsection{SN 2015ca}
\begin{figure*}
\centering
	\begin{tabular}{ccccc}
        \includegraphics[width=0.45\textwidth]{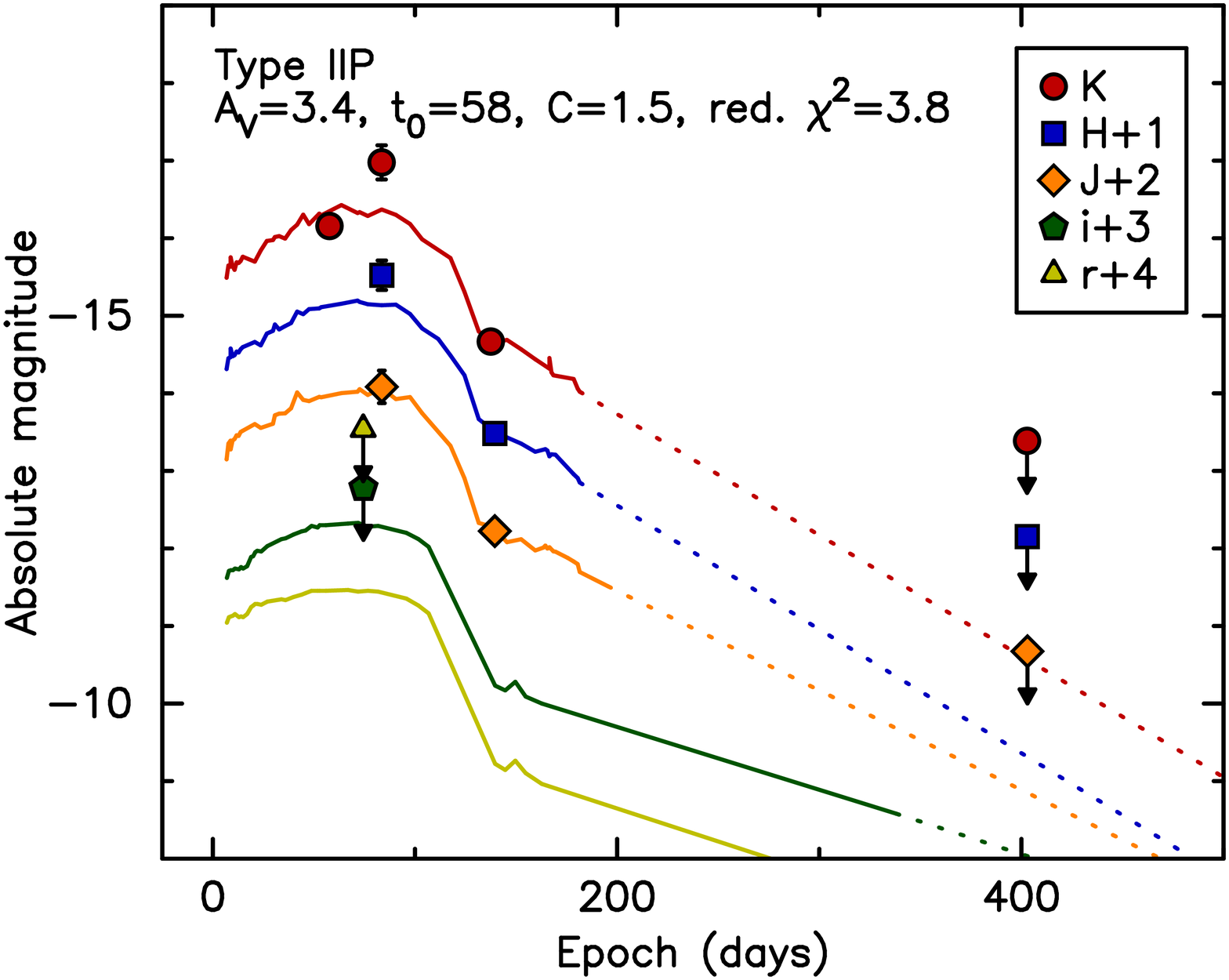} &
        \includegraphics[width=0.45\textwidth]{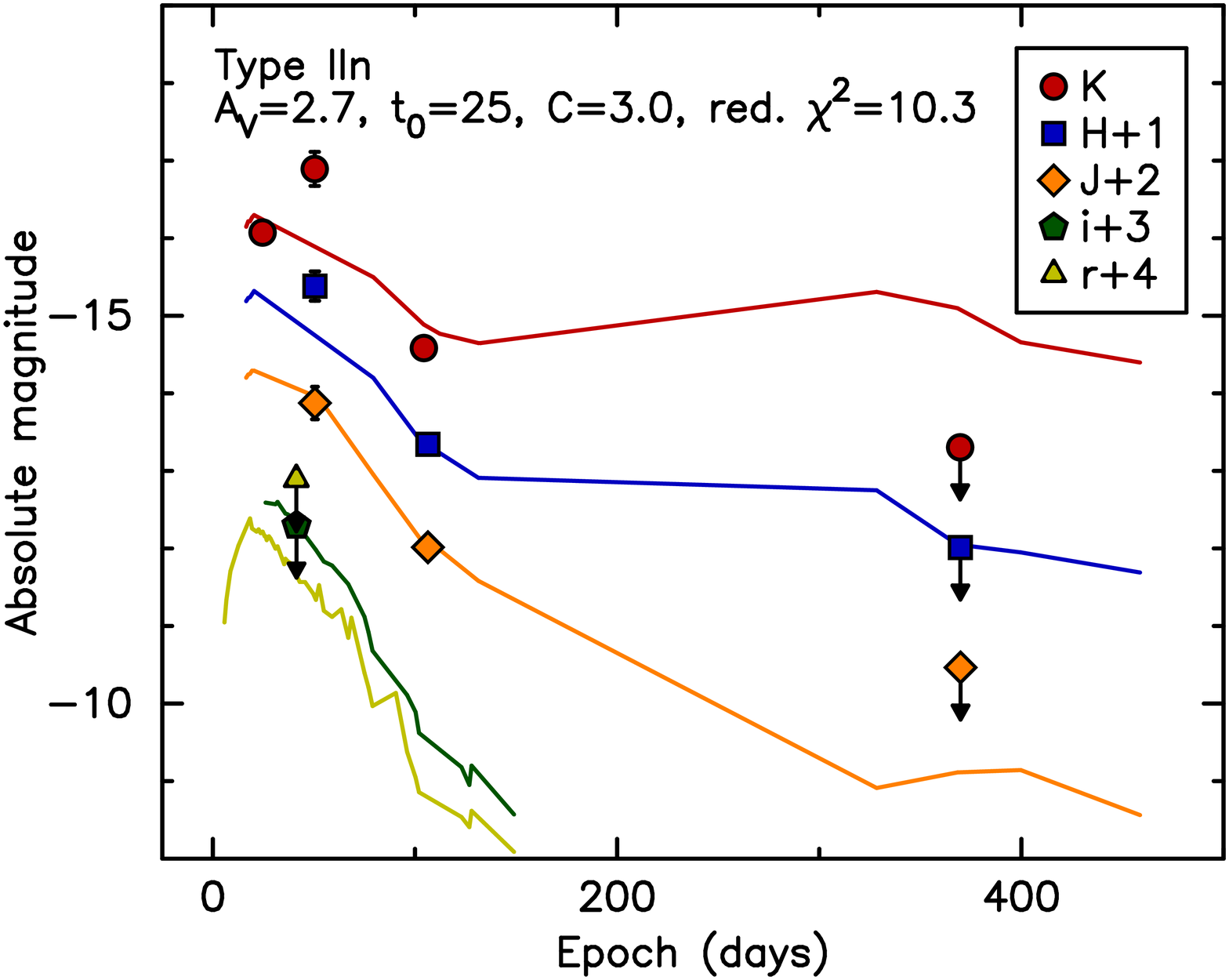} \\
        \includegraphics[width=0.45\textwidth]{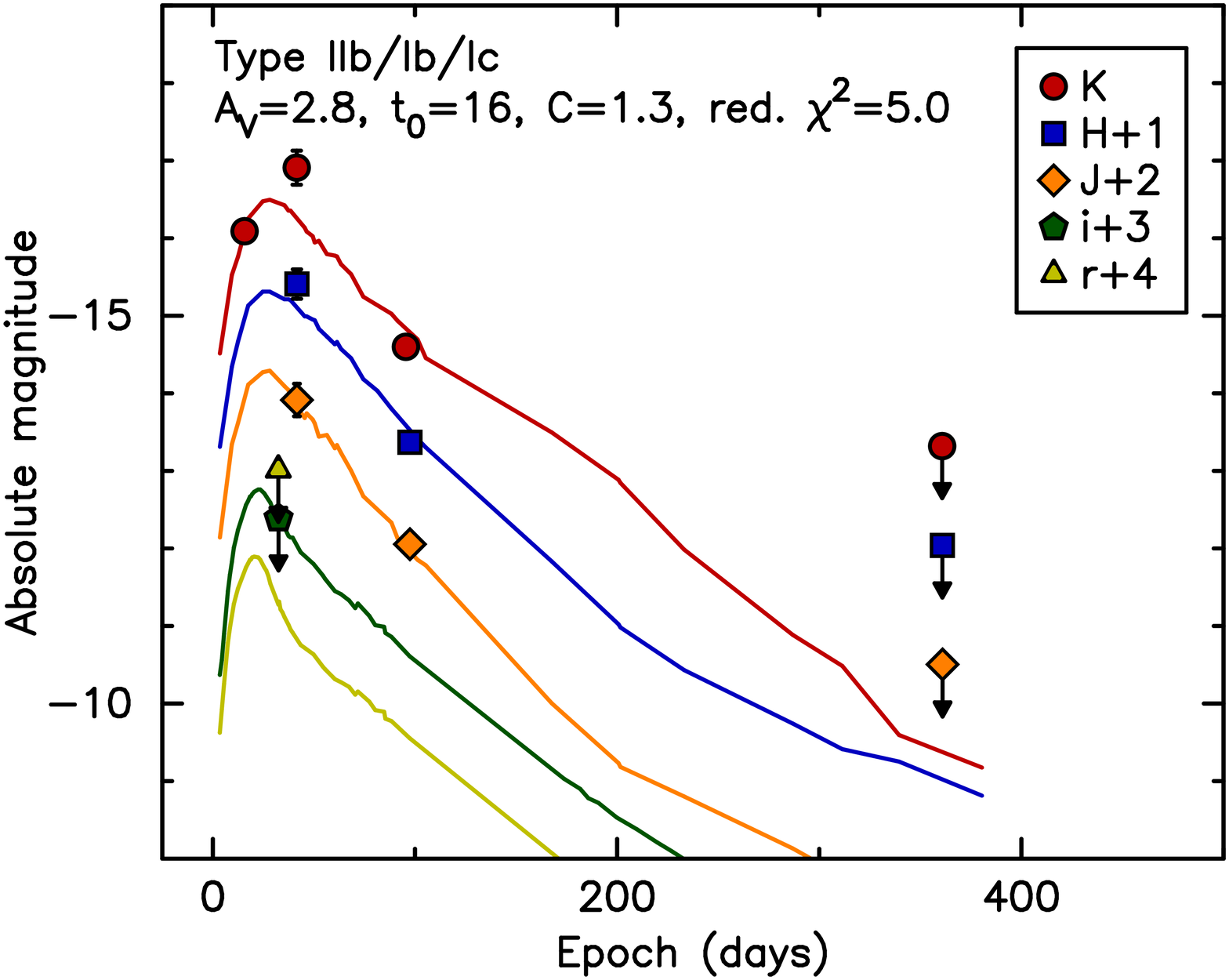} &
        \includegraphics[width=0.45\textwidth]{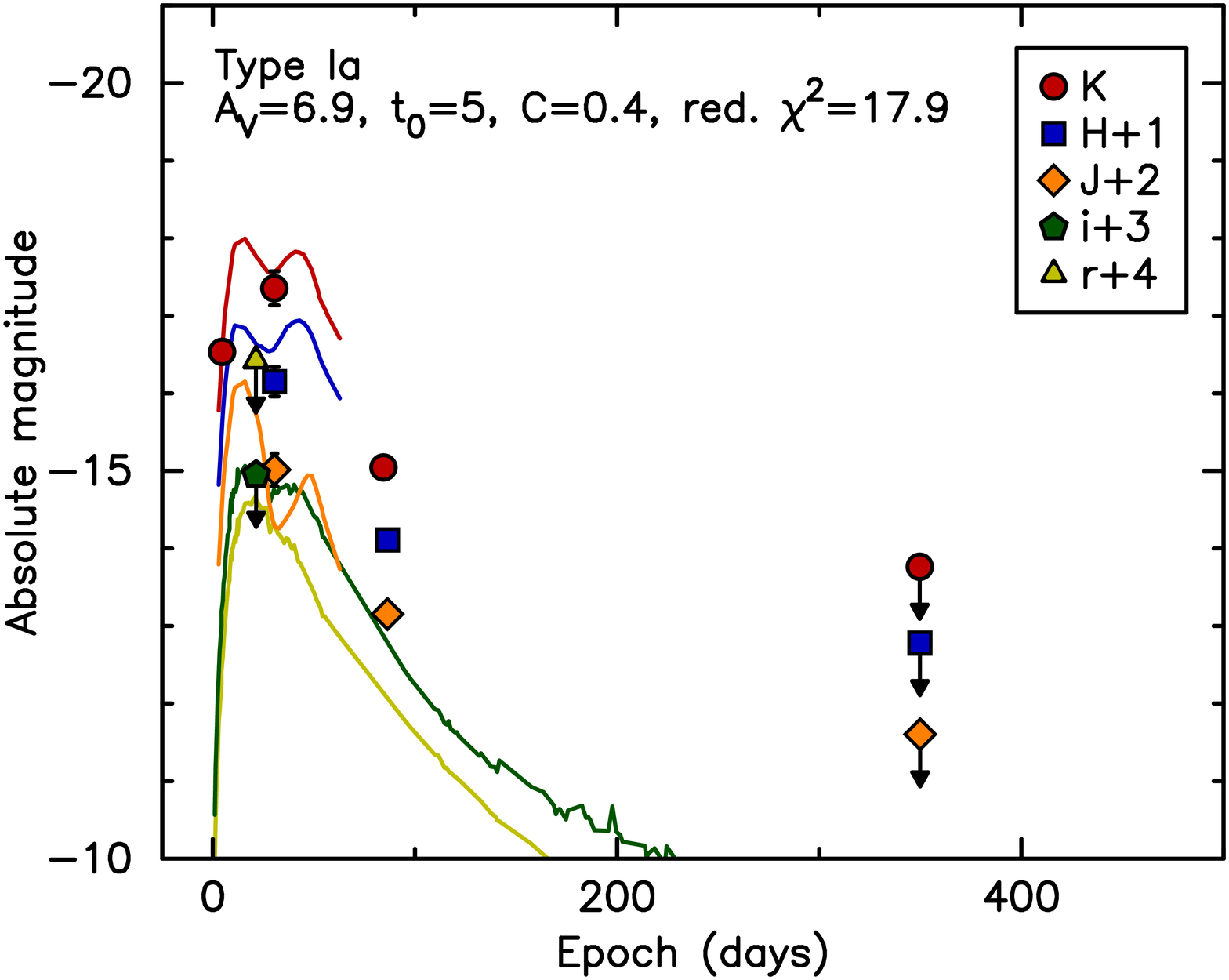} \\
	\end{tabular}
    \caption{Template light curve fits for SN 2015ca in NGC 3110 are shown: Type IIP fitted in the plateau phase, Type IIn, Type IIb/Ib/Ic and Type Ia}
\label{fig:2015ca_fits}
\end{figure*}

SN 2015ca is best fitted by the Type IIP template, with $\tilde{\chi}^2=3.8$, but the stripped envelope template fits the data too, with $\tilde{\chi}^2=5.0$; see Fig.~\ref{fig:2015ca_fits} and Table~\ref{tab:template}. In both cases the fit requires moderate extinction ($A_V=3.4^{+1.0}_{-1.7}$ and $A_V=2.8^{+0.3}_{-0.4}$, respectively) and a magnitude shift of $\sim$1.5. However neither template fits well the brightest epoch from 2015 April 6 with the NOT. The Type IIn template is poorly fitted by the data, and seems inconsistent with the \textit{JHK} upper limits of the final epoch. We also fitted a Type Ia light curve to the data based on the light curve of SN 2011fe \citep{matheson2012,zhang2016}, because SN 2015ca exploded in an isolated location seemingly far from recent SF. With a $\tilde{\chi}^2=17.9$ the Type Ia fit was clearly inferior to the core collapse scenarios. The Type IIn, stripped envelope, and Type Ia fits have been forced to take into account the \emph{i'}-band limit from the NOT, i.e. the \emph{i'}-band curve is basically matching with the i-band limit. The Type IIP fit was consistent with the limit based on the detections.

The follow up at radio wavelengths with the VLA and eEVN did not show any detections of SN 2015ca. For the two likely scenarios the radio luminosity for Type IIP SNe typically peaks between $\sim$20-70 days post explosion and are much fainter than stripped envelope SNe, which peak between 10-150 days post explosion \citep{romero2014}. In the case of the IIP fit, the VLA non-detection would have been $86^{+14}_{-5}$ days post explosion, meaning the relatively faint radio signature of the SN would already have been on the decline. For the stripped envelope fit, the observation occurred $44^{+6}_{-2}$ days post explosion, which coincides more closely with radio peak luminosities for the type. For instance for SN 2011dh, the prototypical SN used for the stripped envelope template, the observation would have been at peak ($\sim$40 days) with a luminosity forty times brighter than SN 1999em, the SN used for the Type IIP template ($\sim$$8\times 10^{26}\ \mathrm{erg}\ \mathrm{s}^{-1}\ \mathrm{Hz}^{-1}$ versus $\sim$$2\times 10^{25}\ \mathrm{erg}\ \mathrm{s}^{-1}\ \mathrm{Hz}^{-1}$, respectively). However, as the SN is coincident with strong host galaxy signal, it is not possible to exclude certain SN types definitively based on the VLA observation.

The eEVN non-detection of SN 2015ca at 5 GHz with a 3$\sigma$ upper limit of 60 $\mu\mathrm{Jy}\,\mathrm{beam}^{-1}$ (<$4\times 10^{26}\ \mathrm{erg}\ \mathrm{s}^{-1}\ \mathrm{Hz}^{-1}$) occurred 60 days post discovery. This would have been $118^{+14}_{-5}$ days for the Type IIP fit and $76^{+6}_{-2}$ days post explosion for the stripped envelope fit. This means a Type IIP SN would have been well past peak luminosity and at any point along the light curve below the detection limit. On the other hand a stripped envelope SN would have been much closer in time to peak with a luminosity well above the upper limit, This is shown in Fig. 4 of \citet{romero2014} for the majority of stripped envelope SNe with the exception of more extensively stripped Type Ic SNe, like 2002ap or 2007gr. As such the eEVN upper limit rules out a large fraction of stripped envelope SNe, implying a Type IIP is more likely for SN 2015ca.

\subsection{SN 2015cb}
\begin{figure*}
\centering
	\begin{tabular}{ccccc}
        \includegraphics[width=0.45\textwidth]{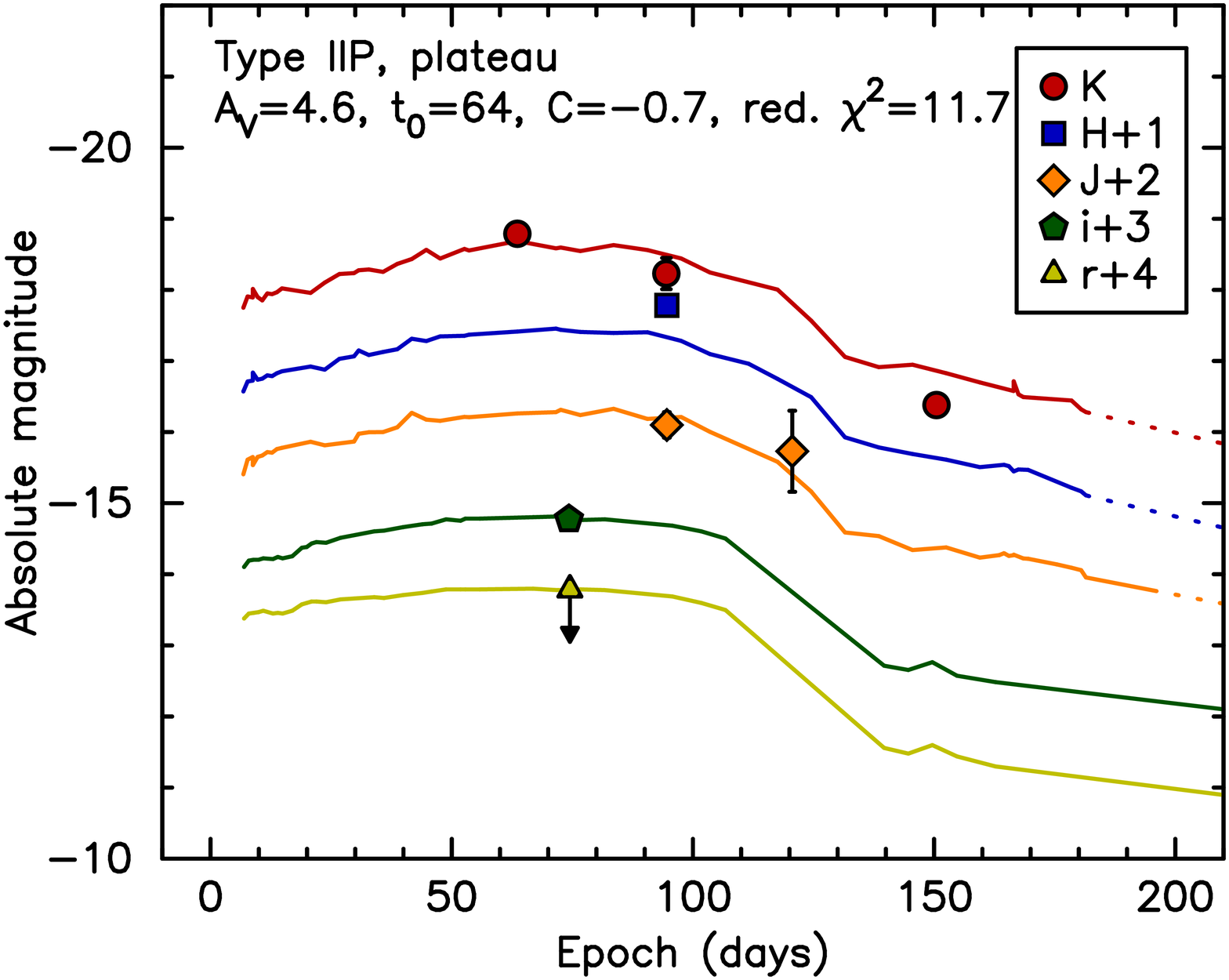} &
       \includegraphics[width=0.45\textwidth]{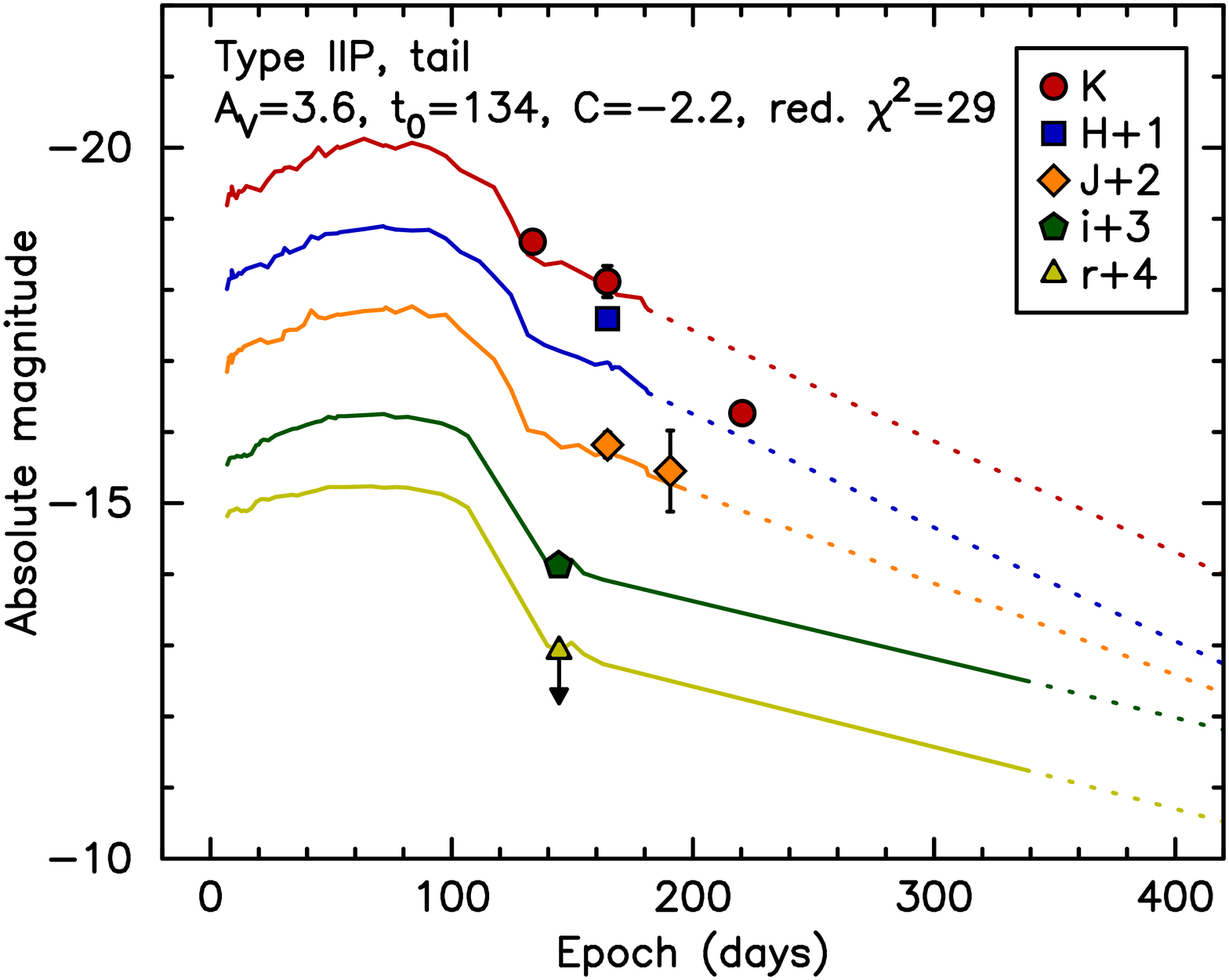} \\
        \includegraphics[width=0.45\textwidth]{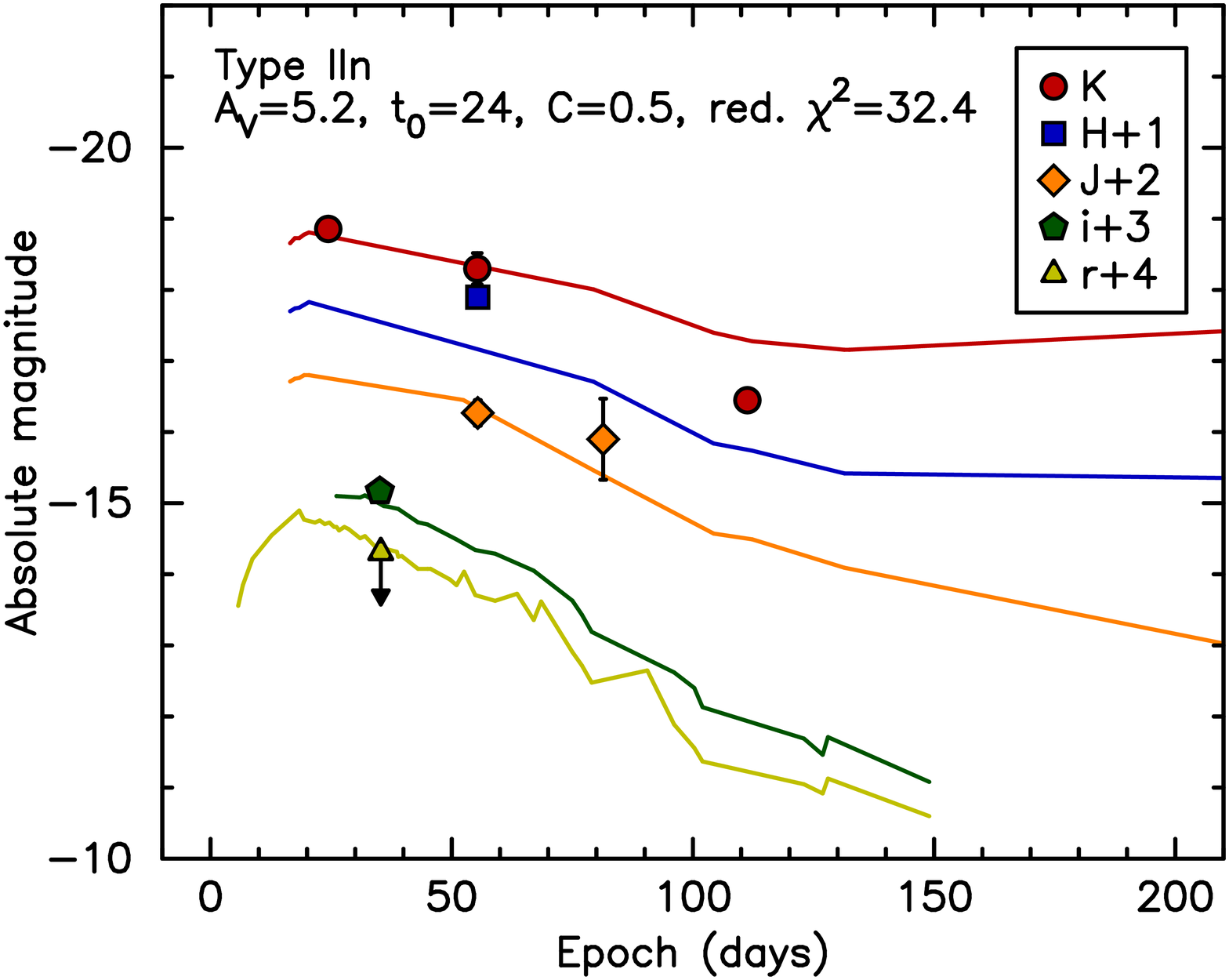} &
        \includegraphics[width=0.45\textwidth]{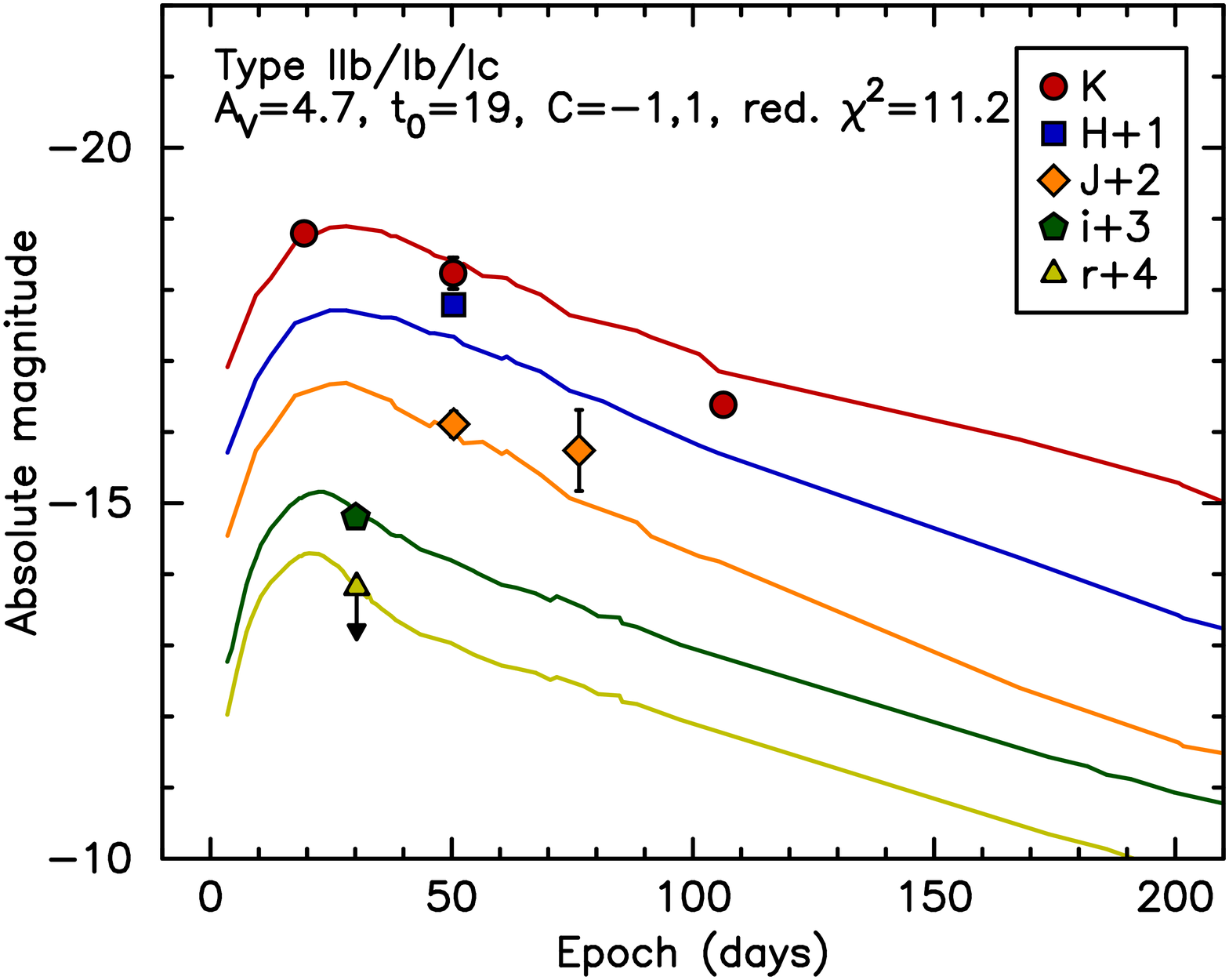} \\
	\end{tabular}
    \caption{Template light curve fits for SN 2015cb in IRAS 17138-1017 are shown: Type IIP fitted in the plateau and tail phase, Type IIn, and Type IIb/Ib/Ic}
\label{fig:2015cb_fits}
\end{figure*}

The SN 2015cb data is best fitted by the templates of a Type IIP caught in the plateau phase or a stripped envelope SN, with $\tilde{\chi}^2=11.7$ and $\tilde{\chi}^2=11.2$, respectively; see Fig.~\ref{fig:2015cb_fits} and Table~\ref{tab:template}. In both cases a slightly more luminous than average SN has been observed (magnitude shifts $C=-0.7^{+0.1}_{-0.3}$ and $C=-1.1^{+0.1}_{-0.2}$, respectively) with a moderate extinction of $\sim$4.5 magnitudes in \textit{V}. To take into account the \emph{r'}-band limit, all template fits were required for the limit to at least match the \emph{r'}-band template curve. Similar to SN 2015ca, the lack of a radio detection with the VLA a month after discovery favours a Type IIP scenario, but the presence of significant contamination around the SN location prevents obtaining a strong upper limit to conclusively exclude a stripped envelope SN. 

\subsection{AT 2015cf}
\begin{figure*}
\centering
	\begin{tabular}{ccccc}
        \includegraphics[width=0.45\textwidth]{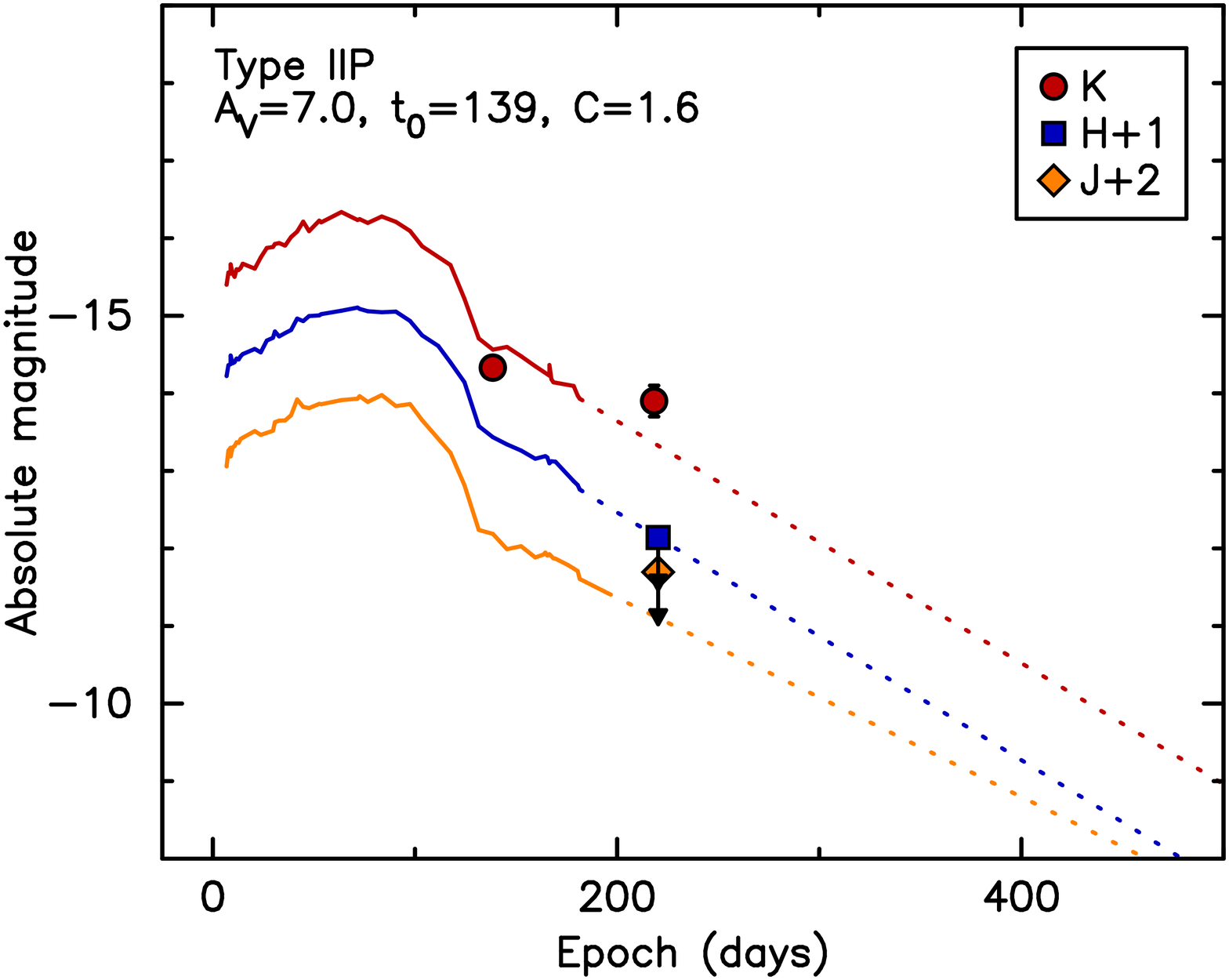} &
       \includegraphics[width=0.45\textwidth]{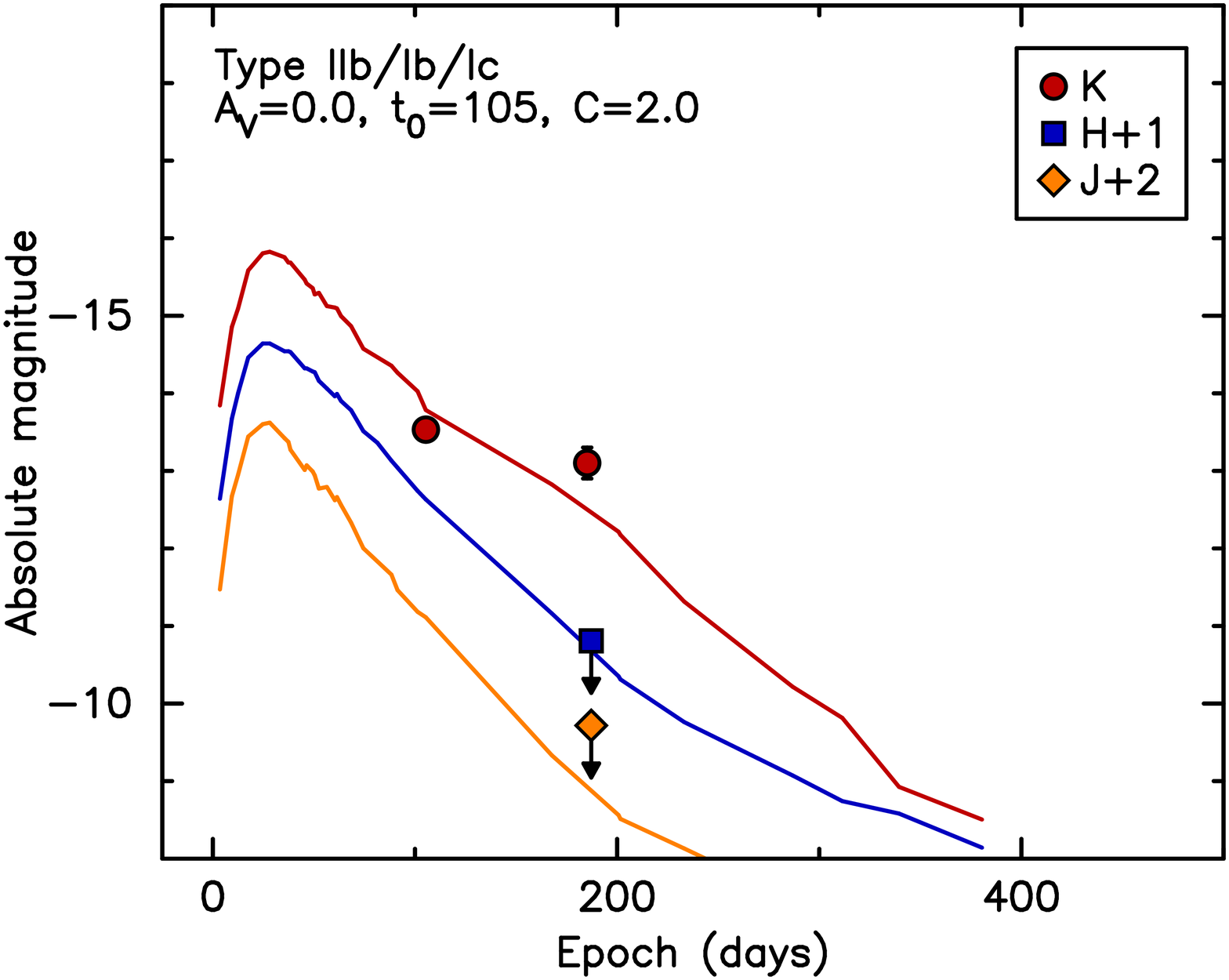} \\
	\end{tabular}
    \caption{Template light curve fits for AT 2015cf in NGC 3110 are shown: Type IIP and Type IIb/Ib/Ic}
\label{fig:2015cf_fits}
\end{figure*}

If AT 2015cf were a young SN, this would yield an unlikely combination of extremely high line-of-sight extinction and/or intrinsically very sub-luminous SN. Furthermore, the \textit{K}$_s$-band decline rate of AT 2015cf from 2015 March 11 until 2015 May 29 (0.5 $\pm$ 0.4 mag / 100 days) is within errors roughly consistent with the theoretical $^{56}$Co decay rate with complete $\gamma$-ray trapping. However, typically H-poor SNe decline more rapidly in the tail phase. Nonetheless, to estimate parameter limits, we consider both the Type IIP and IIb/Ib/Ic template options, with the former providing a somewhat better fit due to the aforementioned decline rate, see Table \ref{tab:template} and Fig. \ref{fig:2015cf_fits}. The NOT observations in \emph{JHKs} from 2015 April 6 cover the site of AT 2015cf, but due to the poorer data quality (FWHM $\sim$1.0\arcsec) and the small separation from SN 2015ca of 0.7\arcsec, the presence of the residual of SN 2015ca in the subtracted image prevents us from obtaining any meaningful upper limits for AT 2015cf. The NOT optical upper limits from 2015 March 27 are the same as obtained for SN 2015ca, but the AT 2015cf data are fitted to a much fainter stage in the templates which means the optical upper limits do not provide useful constraints to the template fits. As such we have opted not to include the NOT upper limits in Table \ref{tab:template} and Fig. \ref{fig:2015cf_fits}. The eEVN non-detection of SN 2015ca does not cover the position of AT 2015cf and cannot be used to constrain the SN type. We conclude that the observations of AT 2015cf are most consistent with an old, possibly H-rich, CCSN.

\subsection{CCSN rate of NGC 3110}
Table \ref{tab:sed} shows for NGC 3110 the SF rate, starburst age, CCSN rate and the origin of its bolometric luminosity. This is based on modelling the multi-wavelength SED of NGC 3110, using data points available from the literature ranging from optical to submillimetre photometry \citep{u2012}, by combining libraries of starburst, AGN torus and disk component models. For more details see \citet{herrero2017} and references therein, with the difference that a spheroidal/cirrus component was fitted instead of a disk (Efstathiou et al, \emph{in preparation}). 
\begin{figure}
	\includegraphics[width=\columnwidth]{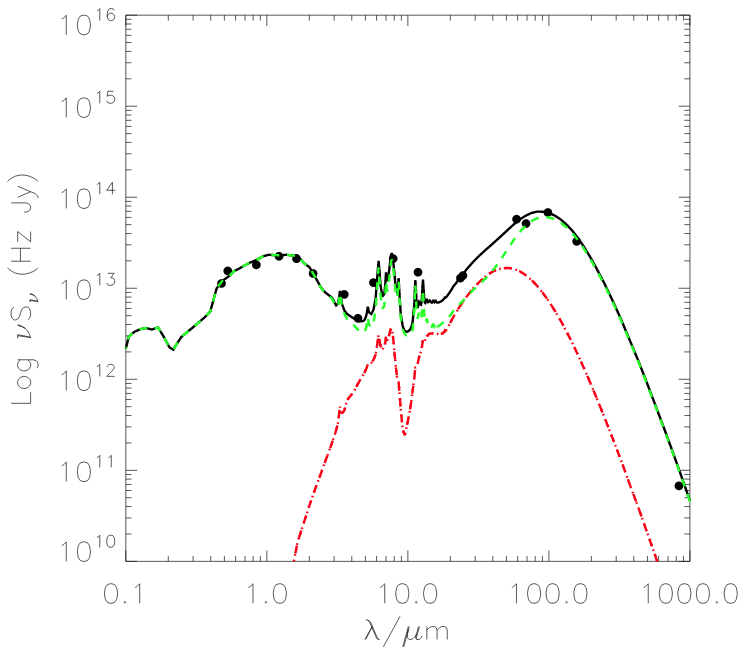}
    \caption{Best SED fitting model of NGC 3110, with the data points indicated by the dots, starburst contribution by the red line and disk component by the green line. No AGN contribution was required to fit the data. The total fit is shown by the black line.}
    \label{fig:sed}
\end{figure}

We found that the SED of the host galaxy NGC 3110 was best fitted by a disk model with a minor starburst component. Table~\ref{tab:sed} shows the best fit parameters and Fig.~\ref{fig:sed} the best fitting model. This means the dominant contributor to the galaxy's total luminosity is not the starburst luminosity, but rather the luminosity of the disk component, which is in strong contrast with the other two SN hosts IRAS 18293-3413 and IRAS 17138-1017 \citep{herrero2017}. This results in a CCSN rate for NGC 3110 which is significantly lower ($0.14^{+0.01}_{-0.01}~\mathrm{SN~yr}^{-1}$) than the expectation from the LIRG's IR luminosity, which would be $\sim0.6~\mathrm{SN~yr}^{-1}$ when applying Eq. \ref{eq:snr}. The reason for this is that as a disk-dominated object, most of the IR luminosity in this galaxy is due to reprocessed radiation from old stars which are not massive enough to explode as SNe. 

Finding two concurrent CCSNe in a galaxy with a CCSN rate of $0.14~\mathrm{yr}^{-1}$ is peculiar, but not impossible. If we assume a Poisson distribution and that SNe are detectable for 3 months, as has been the case for the discoveries in this paper, the probability of discovering two CCSNe in any of the four epochs of NGC 3110 is 4\% and 0.2\% for an expected yearly SN rate of 0.6 and 0.14, respectively. If we assume the same rates for our whole sample (36 epochs in total), this increases to 30.8\% and 1.8\%, respectively.

The CCSN nature of SN 2015ca is well established. The CCSN nature of AT 2015cf remains poorly constrained, but other plausible scenarios with sufficiently high rates are not known to the authors. One possible explanation for this discrepancy is that the starburst age is significantly underestimated. A higher starburst age would give a higher CCSN rate. One important difference between the model fit for NGC 3110 and the other LIRGs studied by \citet{herrero2017} is that in the case of NGC 3110 we lack Spitzer IRS data for the whole galaxy (because of its large angular size) which would constrain our model further.

\subsection{Detection efficiency}
To evaluate how effective our detection method was at recovering SNe, we calculated the detection efficiency in the data by simulating artificial sources in an epoch typical for our sample. We chose to simulate SNe in the GeMS/GSAOI epoch from 2013 April 21 of IRAS 18293-3413, as it is a fairly typical LIRG at a typical distance within our sample and host to SUNBIRD discovery SN 2013if. SNe were simulated in the central region of the galaxy, defined as containing 80\% of the galaxy light. We constructed a model PSF from three bright field stars, placed it in a random position in this region and then performed a subtraction as described in Section \ref{sec:image_sub}. To recover the source, aperture photometry was taken at the simulation location and a 3x3 grid of apertures around it, separated by twice the FWHM. The source was considered recovered when the source flux exceeded three times the standard deviation of the counts recorded in the surrounding aperture grid. The results were split into three regions: the central 100 pc (<0.25\arcsec), where the residuals from subtraction are very large, a wider nuclear region of 300 pc omitting the central 100 pc (0.25\arcsec\ - 0.75\arcsec), and the remaining area which extended to 600-800 pc (0.5\arcsec\ - 2\arcsec).

This was repeated for a range of magnitudes and an S-curve \citep[e.g.][]{dahlen2008,kankare2012} was fitted to the data. We derive a preliminary 50\% detection efficiency at magnitudes of 16.7, 19.3 and 20.7 for the three regions referred to above, respectively. As expected, our detection efficiency is significantly reduced in the central 100 pc of the galaxy. SN 2013if was detected in this galaxy at magnitude 18.53 at 200 pc from the nucleus, halfway the second region, which agrees well with the derived 50\% detection efficiency for the region. For LIRGs such as this, there is significant variation in the chances of detection, heavily dependent on the local galaxy structure and proximity to the nucleus. Additionally, differences in the data quality between epochs affects detection efficiency. This is reflected by the non-detection of SN 2013if in \emph{J}-band (where AO-performance in near-IR typically is worst) from 2013 June 11, with an upper limit magnitude of 18.7. We will present an expanded approach to the evaluation of this across multiple near-IR AO surveys, including SUNBIRD, in a future publication (Reynolds et al, \emph{in prep}).

\section{Discussion}
\label{sec:discussion}
\begin{table*}
\centering
\begin{tabular}{|l|l|l|l|c|c|l|} \hline
Name&LIRG host&RA&Dec.&Type&Projected&Discovered by \\
&&(J2000)&(J2000)&&distance (kpc)& \\ \hline
SN1968A&	NGC 1275&		03:19:49&		+41:30.2&		I&		12& Lovas\\
SN1983V&	NGC 1365&		03:33:31.63&	-36:08:55.0&	Ic&		7& Evans et al.\\
SN1993G&	NGC 3690&		11:28:33.43&	+58:33:31.0&	II&		3.7& Treffers et al.\\
SN1996D&	NGC 1614&		04:34:00.30&	-08:34:44.0&	Ic&		2.1& Drissen et al.\\
SN1998T&	NGC 3690&		11:28:33.14&	+58:33:44.0&	Ib&		1.5& BAOSS\\
SN1999bx&	NGC 6745&		19:01:41.44&	+40:44:52.3&	II&		5&	LOSS \\
SN1999D&	NGC 3690&		11:28:28.38&	+58:33:39.0&	II&		6&	BAOSS \\
SN1999ec&	IC 2163/NGC 2207&	06:16:16.16&	-21:22:09.8&	Ib&	14&	LOSS \\
SN1999ex&	IC 5179&		22:16:07.27&	-36:50:53.7&	Ib&		6&	Martin et al. \\
SN1999gl&	NGC 317B&		00:57:40.07&	+43:47:35.6&	II&		1.9&	Boles \\
SN2000bg&	NGC 6240&		16:52:58.18&	+02:23:51.5&	IIn&	9&	LOSS \\
SN2000cr&	NGC 5395&		13:58:38.37&	+37:26:12.4&	Ic&		10&	Migliardi, Dimai \\
SN2001du&	NGC 1365&		03:33:29.11&	-36:08:32.5&	II&		7&	Evans \\
SN2001eq&	PGC 70417&		23:04:56.78&	+19:33:04.8&	Ic&		6&	LOTOSS \\
SN2001is&	NGC 1961&		05:42:09.07&	+69:21:54.8&	Ib&		14&	BAOSS \\
SN2003H&	IC 2163/NGC 2207&	06:16:25.68&	-21:22:23.8&Ib Pec&	6&	LOTOSS \\
SN2003hg&	NGC 7771&		23:51:24.13&	+20:06:38.3&	IIP&	3.2&	LOTOSS \\
SN2004bf&	UGC 8739&		13:49:15.45&	+35:15:12.5&	Ic&		8&	LOSS \\
SN2004ed&	NGC 6786&		19:10:53.62&	+73:24:27.6&	II&		5&	Armstrong \\
SN2004gh&	MCG-04-25-06&	10:24:31.60&	-23:33:18.4&	II&		2.0&	LOSS \\
SN2005H&	NGC 838&		02:09:38.52&	-10:08:43.6&	II&		0.4&	LOSS \\
SN2007ch&	NGC 6000&		15:49:47.82&	-29:23:13.7&	II&		3.6&	Monard \\
SN2008fq&	NGC 6907&		20:25:06.19&	-24:48:27.6&	II&		1.4&	LOSS \\
SN2009ap&	ESO 138-G27&	17:26:43.23&	-59:55:57.9&	Ic&		1.2&	Pignata et al. \\
SN2010as&	NGC 6000&		15:49:49.23&	-29:23:09.7&	Ib/c&	0.6&	Maza et al. \\
SN2010bt&	NGC 7130&		21:48:20.22&	-34:57:16.5&	II&		6&	Monard \\
SN2010gg&	ESO 602-G25&	22:31:25.47&	-19:01:54.8&	II&		6&	Pignata et al. \\
SN2010gk&	NGC 5433&		14:02:35.94&	+32:30:30.7&	Ic&		2.0&	LOSS \\
SN2010jp&	IC 2163/NGC 2207&	06:16:30.63&	-21:24:36.3&	II&130&	Maza et al. \\
SN2010O&	NGC 3690&		11:28:33.86&	+58:33:51.6&	Ib&		1.6&	Newton, Puckett \\
SN2012by&	UGC 8335&		13:15:28.90&	+62:07:47.8&	II&		10&	Rich \\
SN2013ai&	IC 2163/NGC 2207&	06:16:18.35&	-21:22:32.9&	II&	9&	Conseil \\
SN2013cc&	NGC 1961&		05:41:58.76&	+69:21:40.9&	II&		19&	Itagaki \\
SN2013dc&	NGC 6240&		16:52:58.97&	+02:24:25.2&	IIP&	11&	Block \\
SN2014dj&	NGC 317B&		00:57:40.18&	+43:47:35.1&	Ic&		1.5&	Rich \\
SN2014eh&	NGC 6907&		20:25:03.86&	-24:49:13.3&	Ic&		14&	LOSS \\
SN2015ae&	NGC 7753&		23:47:06.15&	+29:29:07.4&	II&		6&	Itagaki \\
SN2015U&	NGC 2388&		07:28:53.87&	+33:49:10.6&	Ibn&	1.8&	LOSS \\
PS15aaa&	IC 564&			09:46:20.73&	+03:04:22.1&	II&		3.2&	Pan-STARRS \\
\hline
\end{tabular}
\caption{CCSNe hosted by LIRGs discovered in optical, from cross referencing the IRAS Revised Bright Galaxy Sample with the Open Supernova Catalog, Asiago Supernova Catalog, the Transient Name Server and the ASAS-SN supernova sample.}
\label{tab:optical_lirg_sne}
\end{table*}

\begin{table*}
\centering
\begin{tabular}{|l|l|l|l|c|l|c|c|c|} \hline
Name&		LIRG host&			RA&				Dec.&			Type&	Reference&	Extinction &Projected\\
	&		&					(J2000)&		(J2000)&			&				&	A$_V$ (mag)&distance (kpc)\\ \hline
\multicolumn{8}{|c|}{\bf{Non AO}}\\ \hline
SN1992bu&	NGC 3690&			11:28:31.5&		+58:33:38&		Ib/c?&	\citet{vanburen1994}&&1.3 \\ 
SN1999gw&	UGC 4881&			09:15:54.7&		+44:19:55&		II&		\citet{maiolino2002}&&2.9 	\\ 
SN2001db&	NGC 3256&			10:27:50.4&		-43:54:21&		II&		\citet{maiolino2002}&5.5&1.5 \\ 
SN2005U&	NGC 3690&			11:28:33.22&	+58:33:42.5&	IIb&	\citet{mattila2005u}&&1.4 	\\
SN2005V&	NGC 2146&			06:18:38.28&	+78:21:28.8&	Ib/c&	\citet{mattila2005v}&&0.5 	\\
SN2010hp&	MCG-02-01-52&		00:18:50.01&	-10:21:40.6&	IIP&	\citet{miluzio2013}&0.5&2.1 \\
SN2010P&	NGC 3690&			11:28:31.38&	+58:33:49.3&	Ib/IIb&	\citet{kankare2014}&7&1.2 	\\
PSN2010&	IC 4687&			18:13:40.213&	-57:43:28.00&	IIP?&	\citet{miluzio2013}&0-8&6.5 	\\ 
PSN2011&	IC 1623A&			01:07:46.229&	-17:30:29.48&	Ic?&	\citet{miluzio2013}&0.5&3.2  	\\
SN2011ee&	NGC 7674&			23:27:57.34&	+08:46:38.1&	Ic&		\citet{miluzio2013}&0&8.9		\\
SPIRITS 14buu&	IC 2163/NGC 2207&	06:16:27.2&	-21:22:29.2&	IIP?&	\citet{jencson2016}&1.5&2.0	\\
SPIRITS 15c&	IC 2163/NGC 2207&	06:16:28.49&	-21:22:42.2&	IIb&\citet{jencson2016}&2.2&2.1	\\ \hline
\multicolumn{8}{|c|}{\bf{AO}}\\ \hline
SN2004ip&	IRAS 18293-3413&	18:32:41.15&	-34:11:27.5&	II&		\citet{mattila2007}&5-40&0.5\\
SN2004iq&	IRAS 17138-1017&	17:16:35.90&	-10:20:37.9&	II&		\citet{kankare2008}&0-4&0.66 \\
SN2008cs&	IRAS 17138-1017&	17:16:35.86&	-10:20:43.0&	II&		\citet{kankare2008}&17-19&1.5 	\\
SN2010cu&	IC 883&				13:20:35.36&	+34:08:22.2&	II&		\citet{kankare2012}&0-1&0.18 	\\
SN2011hi&	IC 883&				13:20:35.38&	+34:08:22.23&	II&		\citet{kankare2012}&5-7&0.38 	\\
SN2013if&	IRAS 18293-3413&	18:32:41.10&	-34:11:27.24&	IIP&		This work&0-3&0.2\\
SN2015ca&	NGC 3110&			10:04:01.57&	-06:28:25.48&	IIP&		This work&3&3.5\\
SN2015cb&	IRAS 17138-1017&	17:16:35.84&	-10:20:37.48&	II&		This work&4.5&0.6\\
AT2015cf&	NGC 3110&			10:04:01.53&	-06:28:25.84&	II?&		This work&2-5&3.5\\
\hline
\end{tabular}
\caption{CCSNe discovered in near-IR as found in the literature, divided into discoveries in natural seeing conditions and with the assistance of AO. Space based CCSN discoveries are distributed based on the spatial resolution of the discovery image. This list includes CCSNe without spectral confirmation, and are therefore officially potential SNe (PSN) or astronomical transients (AT). Nuclear offsets of the CCSN discoveries in this work are given with 0.1\arcsec\ precision to reflect the uncertainty of the centre of the hosts nucleus.}
\label{tab:nir_lirg_sne}
\end{table*}


With the detections of SN 2013if, SN 2015ca, SN 2015cb, and AT 2015cf, currently there is a total of 60 CCSN optical and near-IR discoveries in LIRGs reported in the literature since 1968: 39 optical discoveries (see Table~\ref{tab:optical_lirg_sne}) and 21 in the near-IR (see Table~\ref{tab:nir_lirg_sne}). Although the discoveries presented in this paper are a minor contribution to the total sample, they are a significant addition to the five previous CCSN discoveries in crowded and obscured regions using near-IR AO observations. We next consider the usefulness of a starburst CCSN survey in the near-IR with the use of LGSAO as compared to the alternatives. Starting with the most well covered temporal baseline, in total 39 reported CCSNe have been discovered in LIRGs in the optical, with 29 since 2000; see Table~\ref{tab:optical_lirg_sne} for a complete list. This table is a result of cross referencing all galaxies in the IRAS RBGS with L$_{\text{IR}}>10^{11}\text{L}_{\odot}$ (corrected for $H_0$ = 70 km s$^{-1}$) with the most up to date SN catalogues available: Open Supernova Catalog \citep{guillochon2016}; Asiago Supernova Catalog \citep{barbon1999}; Transient Name Server\footnote{\urlwofont{https://wis-tns.weizmann.ac.il/}}; ASAS-SN \citep{shappee2014}. When Eq. \ref{eq:snr} is applied to the same sample of galaxies, the collective expected LIRG CCSN rate would be $\sim$250 yr$^{-1}$. This would amount to $\sim$4000 CCSNe since the start of this century, which is two orders of magnitude larger than the actual observed CCSN rate for LIRGs. We note that Eq. \ref{eq:snr} is based on SF-dominated LIRGs, with negligible AGN contribution to $L_{\text{IR}}$. Furthermore a proper comparison can only be done if one would also know the search characteristics of each survey, such as cadence and magnitude limits, which is beyond the scope of this paper. However it is clear that almost all CCSNe exploding in LIRGs are not being observed optically. This is not surprising as current all-sky wide field SN searches are biased towards isolated and bright SNe, and CCSNe hosted by LIRGs are at relatively large distances (most LIRGs are > 50 Mpc) superimposed on bright background emission and likely affected by dust extinction. For example ASAS-SN, an all sky optical supernova search running since June 2013, has discovered $\sim$500 SNe so far, none of which were located in LIRGs \citep{holoien2017c,holoien2017a,holoien2017b}.

Discoveries in the near-IR add another 21 CCSNe to the sample, where we include SNe that were typed based on photometric data. Throughout this discussion we treat AT 2015cf as a real CCSN, but the overall conclusions do not change if this were not the case. The discoveries summarised in Table~\ref{tab:nir_lirg_sne} are all a result of SN searches targeting starburst galaxies and include this work. The table is split into CCSN discoveries in natural seeing conditions (non-AO) and discoveries assisted by AO. As this is primarily a division in spatial resolution, the space-based SPIRITS CCSN discoveries with Spitzer/IRAC (1.5\arcsec\ diffraction limit) have been included in the non-AO sample. The table shows a clear distinction: all but one of the near-IR CCSNe discovered in natural seeing conditions occurred outside the hosts' inner kpc. In contrast, the AO-assisted discoveries have primarily occurred within the inner kpc, with 6 out of 9 discoveries with nuclear offsets <1 kpc. 

\begin{figure}
	\includegraphics[width=\columnwidth]{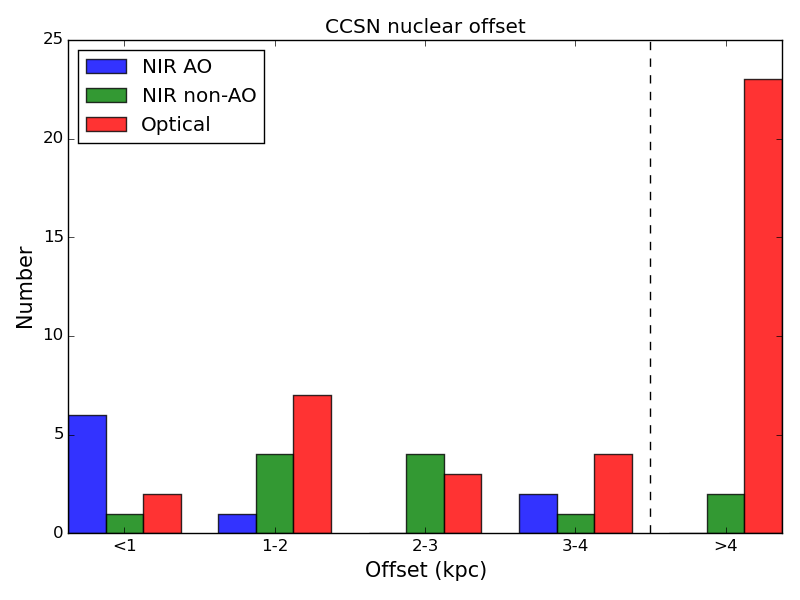}
    \caption{Nuclear offset distribution for all CCSNe discovered in LIRGs. In blue are shown the discoveries in the near-IR with AO instruments, in green the near-IR CCSN discoveries in natural seeing, and in red CCSN discoveries in the optical. The dotted line at 4 kpc marks the limit of the smallest FOV of the instrument used for the AO CCSN discoveries, ALTAIR/NIRI on Gemini North, at a typical distance for LIRGs with SN discoveries of 75 Mpc.}
    \label{fig:distr}
\end{figure}

When the optical CCSN discoveries are added, the benefit of AO to CCSN searches in starbursts becomes even more evident, see Fig.~\ref{fig:distr}. This figure shows the (projected) nuclear offsets of all 60 CCSN discoveries in LIRGs and it is clear that the near-IR programs using AO have been much more effective in finding nuclear CCSNe than both non-AO near-IR and optical programs, even when we disregard discoveries with offsets > 4 kpc. This cutoff is applied to account for selection effects due to differences in FOV. The FOV of AO-imagers is limited and 4 kpc corresponds to 11\arcsec\ at a typical distance of 75 Mpc of a LIRG hosting a discovered SN. This is the limit of the CCSN search radius of ALTAIR/NIRI on Gemini North, the smallest FOV (22\arcsec) of the AO instruments with which CCSNe have been discovered in LIRGs. As such, this should be treated as a lower limit, since for example the FOV of GeMS/GSAOI is 85\arcsec. 


The plotted numbers in Fig. \ref{fig:distr} have not been normalized to account for coverage, so directly comparing the number of discoveries between optical, near-IR and near-IR AO in each radial bin is not appropriate. Normalization would require knowledge of how often LIRGs have been observed in the optical and the near-IR to this day, which is beyond the scope of this paper. However, we can make assumptions and get an order of magnitude estimate of the historical coverages of the three different distributions. For simplicity's sake we will express coverage in terms of number of epochs.

In the near-IR we assume AO-assisted observations of LIRGs are dominated by targeted SN searches, and we know the coverage of these programs. Across \cite{mattila2007} with VLT/NACO, \cite{kankare2008,kankare2012} with ALTAIR/NIRI, and this work with GeMS/GSAOI there have been $\sim$180 epochs in near-IR with AO. If we add the 17 NICMOS epochs from \cite{cresci2007}, this amounts to a total of $\sim$200 epochs of high spatial resolution (AO) imaging of LIRGs.


Individual near-IR non-AO SN programs such as \cite{mannucci2003} (234 epochs) and \cite{miluzio2013} (210 epochs) already exceed the AO sum total. Other programs have targeted starburst galaxies, including LIRGs, such as \cite{grossan1999} ($\sim$500 epochs), \cite{mattila2004} (120 epochs) and the ongoing SPIRITS survey \citep{jencson2016}. Just based on near-IR observations from targeted SN surveys, the coverage of non-AO near-IR observations of LIRGs is a factor of five higher than in the near-IR with AO. Based on this we assume the total coverage in near-IR non-AO is at least an order of magnitude larger than in AO.

By virtue of optical surveys having a larger FOV and greater availability, the coverage of LIRGs in the optical extends over a longer time baseline with higher cadence than in the near-IR. For example the Lick Observatory Supernova Survey (LOSS, \citealt{leaman2011}) observed 14882 galaxies over the course of 11 years, with an average of 150 observations per galaxy. As this sample included $\sim$100 LIRGs, this amounts to 15000 epochs of LIRGs in the optical for LOSS alone. If we also account for optical widefield surveys such as ASAS-SN \citep{shappee2014}, PanSTARRS1 \citep{chambers2016} and CRTS \citep{drake2009} that observe large swaths of the sky on a regular basis, it is safe to say the coverage of LIRG observations in the optical is again at least an order of magnitude higher than in the near-IR.

Based on these relative coverages and the numbers shown in Fig. \ref{fig:distr}, we can conclude the following: First, despite a massive difference in coverage, since 2000 there have been almost as many CCSN discoveries in LIRGs in the near-IR (21) as there have been in the optical (29), which implies SN discovery in LIRGs is substantially more efficient in the near-IR. Second, despite a significant difference in coverage, in the near-IR there have been almost as many CCSN discoveries from AO programs (9) as from non-AO programs (13), with the AO discoveries weighted towards the nucleus regions. The simplest explanation for these conclusions is the combination of reduced dust extinction in the near-IR, and the improved spatial resolution for AO-assisted imaging at these wavelengths, both of which provide enhanced sensitivity to CCSNe detection in these dusty and compact star forming objects.

The CCSN radial distribution as seen in near-IR AO studies agrees well with work on the spatial distribution of CCSNe and CCSN remnants in the nuclear regions of three starburst galaxies and a sample of spiral galaxies using high-angular resolution ($\lesssim$ 0.1\arcsec) radio VLBI observations studied by \cite{herrero2012}. Here it was found that the SN radial distribution in the LIRGs Arp 220 and Arp 299 was centrally peaked with a very steep SN surface number density profile when compared to the SN radial distribution in regular spiral galaxies from \cite{hakobyan2009}. The SNe and SN remnants in \cite{herrero2012} have radial distances almost exclusively smaller than the most centrally located SN in Table \ref{tab:nir_lirg_sne}, which is not surprising given the superior spatial resolution, but limited FOV, of VLBI.

\section{Conclusions}
\label{sec:conclusions}
We have introduced project SUNBIRD, a systematic search for CCSNe in nearby LIRGs using AO imaging at near-IR wavelengths, where we aim to characterize the population of CCSNe hidden in obscured and crowded regions of star forming galaxies. We have presented the first results of this project: so far we have covered a sample of 13 LIRGs with GeMS/GSAOI on the Gemini South telescope and discovered three photometrically confirmed CCSNe and one CCSN candidate. Two of the three CCSNe occurred close to the nucleus of their host galaxies. SN 2013if had a projected distance from the nucleus as small as 200 pc, which makes it the second most nuclear CCSN discovery in a LIRG to date in the optical and near-IR after SN 2010cu \citep{kankare2012}. 

To investigate the impact of these new discoveries and the effectiveness of our near-IR AO CCSN search, we gathered from the literature all CCSN discoveries in LIRGs in the optical and the near-IR. This sample consists of 60 events, out of which 9 were discovered using near-IR AO, including this project. We show a clear distinction in nuclear offset distribution between these 9 events and the other optical and near-IR discoveries. Almost all CCSN discoveries from AO programs occurred in the inner kpc of their hosts, but only a fraction of the non-AO near-IR and optical CCSNe had offsets smaller than 1 kpc. This tells us that our approach is singularly effective in uncovering CCSNe in nuclear regions, and crucial in characterizing this population of CSSNe that will remain invisible through other means.

\section*{Acknowledgements}
We would like to thank the anonymous referee for constructive comments. This publication is based on observations obtained at the Gemini Observatory, which is operated by the Association of Universities for Research in Astronomy, Inc., under a cooperative agreement with the NSF on behalf of the Gemini partnership: the National Science Foundation (United States), the National Research Council (Canada), CONICYT (Chile), Ministerio de Ciencia, Tecnolog\'{i}a e Innovaci\'{o}n Productiva (Argentina), and Minist\'{e}rio da Ci\^{e}ncia, Tecnologia e Inova\c{c}\~{a}o (Brazil). The relevant program codes are: GS-2012B-SV-407 (PI: S. Ryder), GS-2013A-Q-9 (PI: S. Ryder/F. Bauer), GS-2015A-C-2, (PI: S. Sweet/R. Sharp), GS-2015A-Q-6 (PI: S. Ryder), GS-2015A-Q-7 (PI: S. Ryder), GN-2015A-DD-4 (PI: S. Ryder) and GS-2016A-C-1 (PI: E. Kool).

This publication is based on observations made with the Nordic Optical Telescope, operated by the Nordic Optical Telescope Scientific Association at the Observatorio del Roque de los Muchachos, La Palma, Spain, of the Instituto de Astrofisica de Canarias.

This publication is based on observations made with ESO Telescopes at the La Silla Paranal Observatory under programme ID's 073.D-0406 (PI: P. V\"ais\"anen), 086.B-0901 (PI: A. Escala) and 089.D-0847 (PI: S. Mattila), and on data obtained from the ESO Science Archive Facility under request numbers eckool194907, -196798 and -226876.

The National Radio Astronomy Observatory is a facility of the National Science Foundation operated under cooperative agreement by Associated Universities, Inc.

The European VLBI Network is a joint facility of independent European, African, Asian, and North American radio astronomy institutes. Scientific results from data presented in this publication are derived from the following EVN project code(s): RSP13 (P.I. P\'erez-Torres)

RMcD is the recipient of an Australian Research Council Future Fellowship (project number FT150100333).

ECK is grateful for financial support provided by the International Macquarie University Research Excellence Scholarship and the Australian Astronomical Observatory (AAO) through the AAO PhD Scholarship Scheme.

MAPT, RHI, and AA acknowledge support by the Spanish MINECO through grants AYA2012-38491-C02-02 and AYA2015-63939-C2-1-P, co-funded with FEDER funds.

FEB acknowledges support from CONICYT-Chile Basal-CATA PFB-06/2007 and FONDECYT Regular 1141218. CRC acknowledges support from CONICYT through FONDECYT grant 3150238. FEB and CRC acknowledge support from the Ministry of Economy, Development, and Tourism's Millennium Science Initiative through grant IC120009, awarded to The Millennium Institute of Astrophysics, MAS. 




\bibliographystyle{mnras}
\bibliography{SUNBIRD_paper_mnras} 



\newpage
\appendix
\section{Data reduction with \textsc{theli}}
\label{sec:theli_astro}
Described here are optimisations or adaptations of the procedures from \citet{schirmer2013} and \citet{schirmer2015}. The focus is on astrometric calibration of images with low number source density, required for correcting the distortion pattern present in GeMS/GSAOI data.

\subsection{Calibration and background modelling}
The science exposures were divided by a master flat combined from typically 10 dome flats, and the background removed by running a two-pass background model. In short, the first pass is a simple background model subtraction of a median combination of all exposures without masking to remove the bulk of the signal, whereas in the second pass objects are masked by applying SExtractor \citep{bertin1996} for source identification. A static background model was used if the exposure sequence did not last longer than 1 hour. In a few cases a dynamic model was warranted using a running median of 6 images. In contrast to \citet{schirmer2015} no \emph{mask expansion factor} or \emph{collapse correction} was deemed necessary for this work.
\subsection{Astrometric calibration}
The astrometric calibration of individual dithered images prior to coaddition in \textsc{theli} enables the distortion correction of GeMS/GSAOI data, but also requires the most attention to properly process the data. Because of the low number source density, no single combination of parameter settings for the packages that \textsc{theli} employs ({\it Scamp}, \emph{Swarp}: \cite{bertin2002} and SExtractor) proved to be sufficient for all GeMS/GSAOI data sets, some informed adjustments were always required. Most settings are well covered in the \textsc{theli} documentation and aforementioned procedures, but the following adjustments were not obvious and were vital for the astrometric calibration of our GeMS/GSAOI data.

In all of the datasets the comparatively small FOV of GSAOI in combination with sparsely populated fields meant all-sky astrometric reference catalogues were insufficient for astrometric calibration of the individual frames. Instead secondary reference catalogues were first created with \textsc{theli} based on $K_s$-band archival data of ground-based widefield instruments: HAWK-I on the VLT or VIRCAM on the VISTA telescope. These were calibrated against typically 100-1000 2MASS sources in the FOV, which have individual uncertainties of $\sim$100 mas, resulting in astrometric uncertainties of the reference images of $\lesssim$ 10 mas. After successfully constructing a distortion corrected coadded GeMS/GSAOI image, subsequent GeMS/GSAOI epochs were calibrated against a reference catalogue extracted from this image. Depending on field crowding of the astrometric reference image and its spatial resolution, the parameter \textsc{deblend\textunderscore mincont} in \emph{postcoadd.conf.sex} in the \textsc{theli} reduction folder needed to be adjusted (decreased when working with a GeMS/GSAOI reference image).

The WCS header information in our raw GeMS/GSAOI data is generally accurate within $\sim$5\arcsec\ but this proved to be not precise enough in fields with few matching reference sources. In almost all cases it was necessary to update the \textsc{crpix1/2} keywords in the headers to match with the coordinates of the reference catalogue. Then, as described in section 3.5.1 in \citet{schirmer2015}, a refined median estimate of the relative array positions and orientations from all exposures in a night is used (\textsc{mosaic\textunderscore type}\ $=$\ \textsc{fix\textunderscore focalplane}), and as we have adjusted the reference pixel manually no WCS matching is required: \textsc{match}\ $=$\ \textsc{n}.

\textsc{theli} uses \text{SExtractor} to create a source catalogue from the science images which are then fitted to the corresponding sources (within \textsc{crossid\textunderscore radius}) in the reference catalogue to obtain a distortion correction. This process is done for each array separately and the \text{SExtractor} settings might not be appropriate for both the array containing the galaxy (high source number density) and an array covering a sparse field. In the case of NGC~3110 in order to recover all four arrays it was necessary to run the source extraction individually with appropriate settings (most importantly detection thresholds \textsc{detect\textunderscore thresh} and \textsc{detect\textunderscore minarea} and de-blending parameter \textsc{deblend\textunderscore mincont}). This can be achieved by suspending the parallel manager script and the details are described in the \textsc{theli} documentation\footnote{https://astro.uni-bonn.de/$\sim$theli/gui/advancedusage.html}.




\bsp	
\label{lastpage}
\end{document}